\documentclass[12pt,preprint]{aastex}

\newcommand{\Rs}{\ensuremath{R_{\odot}}}
\newcommand{\Ms}{\ensuremath{M_{\odot}}}
\newcommand{\eg}{{\it e.g.}}
\newcommand{\ie}{{\it i.e.}}
\newcommand{\viz}{{\it viz.}}
\slugcomment{To appear in Ap. J.}

\shorttitle{Compact Binary Evolution in Globular Clusters}
\shortauthors{Banerjee and Ghosh}

\begin{document}

%% LaTeX will automatically break titles if they run longer than
%% one line. However, you may use \\ to force a line break if
%% you desire.

\title{Evolution of Compact-Binary Populations\\ 
in Globular Clusters: A Boltzmann Study\\ 
I. The Continuous Limit}

%% Use \author, \affil, and the \and command to format
%% author and affiliation information.
%% Note that \email has replaced the old \authoremail command
%% from AASTeX v4.0. You can use \email to mark an email address
%% anywhere in the paper, not just in the front matter.
%% As in the title, use \\ to force line breaks.

\author{Sambaran Banerjee and 
Pranab Ghosh}
\affil{Department of Astronomy \& Astrophysics \\ Tata Institute of 
Fundamental Research, Mumbai 400 005, India}

%% Mark off your abstract in the ``abstract'' environment. In the manuscript
%% style, abstract will output a Received/Accepted line after the
%% title and affiliation information. No date will appear since the author
%% does not have this information. The dates will be filled in by the
%% editorial office after submission.

\begin{abstract}

We explore a Boltzmann scheme for studying the evolution of compact binary  
populations of globular clusters. We include processes of compact-binary 
formation by tidal capture and exchange encounters, binary destruction by 
dissociation and other mechanisms, and binary hardening by encounters, 
gravitational radiation and magnetic braking, as also the orbital evolution
during mass transfer, following Roche lobe contact. For the encounter 
processes which are stochastic in nature, we study the probabilistic, 
continuous limit in this introductory work, deferring the specific handling 
of the stochastic terms to the next step.  We focus on the evolution of
(a) the number of X-ray sources $N_{XB}$ in globular clusters, and (b) the
orbital-period distribution of the X-ray binaries, as a result of the above 
processes. We investigate the dependence of $N_{XB}$ on two essential 
cluster properties, namely, the star-star and star-binary encounter-rate 
parameters $\Gamma$ and $\gamma$, which we call Verbunt parameters. We 
compare our model results with observation, showing that the model values 
of $N_{XB}$ and their expected scaling with the Verbunt parameters are in 
good agreement with results from recent X-ray observations of Galactic 
globular clusters, encouraging us to build more detailed models. 

\end{abstract}

%% Keywords should appear after the \end{abstract} command. The uncommented
%% example has been keyed in ApJ style. See the instructions to authors
%% for the journal to which you are submitting your paper to determine
%% what keyword punctuation is appropriate.

%% PROVIDE MORE KEYWORDS

%% From the front matter, we move on to the body of the paper.
%% In the first two sections, notice the use of the natbib \citep
%% and \citet commands to identify citations.  The citations are
%% tied to the reference list via symbolic KEYs. The KEY corresponds
%% to the KEY in the \bibitem in the reference list below. We have
%% chosen the first three characters of the first author's name plus
%% the last two numeral of the year of publication as our KEY for
%% each reference.

%% Authors who wish to have the most important objects in their paper
%% linked in the electronic edition to a data center may do so by tagging
%% their objects with \objectname{} or \object{}.  Each macro takes the
%% object name as its required argument. The optional, square-bracket 
%% argument should be used in cases where the data center identification
%% differs from what is to be printed in the paper.  The text appearing 
%% in curly braces is what will appear in print in the published paper. 
%% If the object name is recognized by the data centers, it will be linked
%% in the electronic edition to the object data available at the data centers  
%%
%% Note that for sources with brackets in their names, e.g. [WEG2004] 14h-090,
%% the brackets must be escaped with backslashes when used in the first
%% square-bracket argument, for instance, \object[\[WEG2004\] 14h-090]{90}).
%%  Otherwise, LaTeX will issue an error. 

\section{Introduction}\label{intro}

In this era of high-resolution X-ray observations with \emph{Chandra}
and \emph{XMM-Newton}, studies of compact binaries in globular clusters
have reached an unprecedented level of richness and detail. The numbers
of compact X-ray binaries detected in Galactic globular clusters with 
high central densities are now becoming large enough that diagnostic 
correlations with essential cluster parameters, such as the two-body 
encounter rate $\Gamma$, can be performed \citep{pool03} at a high
level of statistical significance. The results of such observational 
studies are naturally to be compared with those obtained from   
theoretical modeling of binary dynamics in globular clusters, which 
has had a long history, from the pioneering semi-analytic work of the 
1970s \citep{h75}, to the more detailed numerical scattering experiments 
of the 1980s \citep{hb83}, leading to the wealth of detailed numerical 
work of the early- to mid-1990s \citep{ma92,hh2003} using a variety of 
techniques including Fokker-Planck and Monte Carlo approaches, as also 
N-body simulations, and finally to the extensive N-body simulations in 
the latter half of the 1990s using special-purpose computers with 
ultrahigh speeds \citep{mt98,ht2001}. 

The range of problems studied by the above modeling has also been
extensive. From the study and classification of individual scattering
events to the construction of comprehensive fitting formulas for the 
cross-sections of such events \citep{hb83,hhm96}, from the development 
of Fokker-Planck codes to the use of Monte Carlo methods for following
binary distributions in globular clusters \citep{g91,H.et.al92}, and 
from tracking the fate of a relatively modest population of test
binaries against a fixed stellar background to being able to tackle 
similar projects for much larger binary populations with the aid of the 
above special-purpose machines \citep{h.et.al92,m96}, efforts along 
various lines of approach have shed light on the overall phenomenon of 
binary dynamics and evolution in globular clusters from various angles. 
For example, evolutions of the distributions of both \emph{external} and 
\emph{internal} binding energies of the binaries under stellar encounters 
have been studied by several authors, the emphasis usually being on the 
former, and final results on the external binding energy being expressed 
almost universally in terms of their radial positions $r$ inside the 
cluster, which provides an equivalent description 
\citep{H.et.al92,sp93,sp95}.  

In this series of papers, we introduce an alternative method of studying the 
evolution of compact-binary populations in globular clusters, wherein we 
use a Boltzmann description to follow the time-evolution of such populations, 
subject to both (a) those processes which determine compact-binary evolution
in isolation (\ie, outside globular clusters, or, in the ``field'' of the 
host galaxy, so to speak), \eg, angular momentum loss by gravitational 
radiation and magnetic braking, as also orbital evolution due to mass 
transfer, and, (b) those processes which arise from encounters of compact 
binaries with the dense stellar background in globular clusters, \eg, 
collisional hardening \citep{h75,sh79,bg2006}, binary formation through 
tidal capture and exchange processes, and binary destruction. We treat all 
of the above processes simultaneously through a Boltzmann formalism, the aim
being to see their combined effect on the compact-binary population as a 
whole, in particular on the evolution of (a) the total number of X-ray 
binaries as the formation and destruction processes continue to operate, 
and, (b) the orbital-period distribution of the population. We stress at 
the outset that ours is not a Fokker-Planck description but the original 
Boltzmann one, which in principle is capable of handling both the 
\emph{combined} small effects of a large number of frequent, weak,
distant encounters \emph{and} the \emph{individual} large effects of a
small number of rare, strong, close encounters. In our approach, both of 
the above two types of effects are taken into account through 
cross-sections for the relevant processes, as determined from extensive 
previous work on numerical experiments with two-body and three-body 
encounters \citep{hhm96,z297}. As these processes are inherently 
stochastic, a natural question that arises is how they are to be handled 
simultaneously with those which govern the fate of isolated compact 
binaries, and which are inherently continuous. It is essential to 
appreciate the importance of this question, since a simultaneous action 
of the above continuous and stochastic processes is precisely what 
operates on binaries in globular clusters, and so produces the observed 
properties of compact-binary populations in them.    

Our answer to the above question is a step-by-step one, as follows. As
the first step, in this first paper of the series (henceforth Paper I), 
we explore the continuous limit of the above stochastic processes, wherein 
the probability or cross-section of a particular such process happening 
with a given set of input and output variables is treated as a continuous 
function of these variables. This is, of course, a simplification, but it 
serves as a clarification of the average, long-term trends expected in the 
evolution of the binary population. In the next step, in the second paper 
of the series (henceforth Paper II), we treat the stochastic processes as 
stochastic terms in the Boltzmann equation with cross-sections as given in 
Paper I, with the aid of relatively recently-developed methods for solving 
stochastic partial differential equations. The resulting evolutionary 
trends show stochastic behavior, as expected, with fluctuations that vary 
from one particular ``realization'' of the essential processes to another. 
However, the average trends follow the continuous limit computed in Paper 
I, which is as expected, and which shows the relevance of extracting this 
limit. 

In Papers I and II, we model the stellar background provided by the 
globular cluster as a fixed background with given properties, as has been 
widely done in previous works \citep{H.et.al92,z297,sp93,sp95}: 
this amounts to neglecting the back reaction of binary evolution on the 
background, which is reasonable if the main aim is an investigation of 
essential features of binary evolution, as was the case in the above 
previous works, as also in this work. However, the globular-cluster 
background does evolve slowly, passes through the core-collapse phase and 
possible gravothermal oscillations \citep{sb83,g91}, so that it
would be interesting to be able to follow the effects of these on the 
evolution of the compact-binary population. We do this in the third paper
of the series (henceforth Paper III), wherein we adopt previous results on
time-evolution of globular-cluster properties, and study their effects on
the evolution of compact-binary populations, again under the approximation
of neglecting the back reaction of binary evolution on the globular-cluster
background, as above and as appropriate for a first look.      

In our study, we focus primarily on two aspects of the compact-binary 
populations of globular clusters. First, we study how the total number 
$N_{XB}$ of X-ray binaries (henceforth XBs, which are mass-transferring 
compact binaries where the donor is a low-mass ``normal'' star, and the 
accretor is a degenerate star --- a neutron star or a heavy white dwarf) 
in a cluster evolves as the stellar encounter processes proceed.  
Second, we also follow the evolution of the orbital-period ($P$) 
distribution of the pre-X-ray binaries (henceforth PXBs; also see below) 
and XBs, (or, equivalently, the distribution of their orbital radii $a$) 
within the framework of our model. However, we have adopted here only a 
very simple model of orbital evolution of individual binaries in order to 
assess the feasibility of our basic approach to globular-cluster 
environments, as detailed later. Consequently, while the $P$-distribution 
found by us may be roughly applicable to cataclysmic variables (CVs) with 
white-dwarf accretors, it cannot be compared at this stage to that of 
low-mass X-ray binaries (LMXBs) with neutron-star accretors, without including 
the essential stellar evolutionary processes that occur during the PXB
and XB phase. Thus, we record our computed $P$-distribution here only as 
a preliminary indication of the results that emerge naturally from this
line of study at this stage, to be improved upon later. 
     
The basic motivation for our study comes from recent advances in X-ray 
observations of globular clusters, as mentioned above: with sufficient 
numbers of X-ray binaries detected in globular clusters, an understanding 
of how $N_{XB}$ is influenced by essential globular-cluster parameters is 
becoming a central question. With the above goal in mind, we therefore 
explicitly follow the evolution of binaries only in internal binding 
energy (or binary period, or binary separation, which are equivalent 
descriptions if the stellar masses are known) and time, but not of their 
external binding energy (or position inside the globular cluster; see 
above). We emphasize that we do not \emph{neglect} changes in the latter 
in any way, as they are automatically taken care of in the detailed dynamics 
of encounters which are represented by the relevant cross-sections mentioned 
above and elaborated on in the following sections. It is only that we do not 
keep an explicit account of them, as we do not need them for our purposes. 
In other words, we consider a bivariate binary distribution function 
$n(E_{in},t)$, which may be looked upon as the integral of the distribution 
$\rho(E_{ex},E_{in},t)$ over all admissible values of $E_{ex}$, or 
equivalently over all positions $r$ inside the globular cluster 
\citep{H.et.al92,sp93,sp95}. We also emphasize that, by doing so, we do 
\emph{not} implicitly assume any particular correlation, nor a lack thereof, 
between $E_{in}$ and $E_{ex}$ \citep{H.et.al92}: whatever correlations 
result from the dynamics of the encounters will be automatically displayed 
if we follow the evolution in $E_{ex}$ or $r$, which is not of interest to 
us in this particular study. 

Our first results from the above evolutionary scheme show that the total 
number $N_{XB}$ of XBs expected in a globular cluster scales in a
characteristic way with well-known globular cluster parameters $\Gamma$ and
$\gamma$ (which we call Verbunt parameters: see Sec.~\ref{gc}) whose
qualitative nature is rather similar to that found in our earlier ``toy'' 
model \citep{bg2006}, although some details are different. Basically, 
$N_{XB}$ scales with $\Gamma$ --- a measure of the dynamical formation 
rate of compact binaries, and, at a given $\Gamma$,  $N_{XB}$ decreases with 
increasing $\gamma$ at large values of $\gamma$ --- a measure of the rate 
of destruction of these binaries by dynamical processes. These expected 
theoretical trends with the Verbunt parameters compare very well with the 
observed trends in recent data, encouraging us to construct more detailed 
evolutionary schemes.

In Sec.~\ref{model}, we detail our model of compact binary evolution in 
globular clusters, describing, in turn, our handling of globular clusters, 
binary formation, destruction, and hardening processes, our Boltzmann 
scheme for handling population-evolution, and our numerical method. 
In Sec.~\ref{results}, we give our model results on (a) the expected number 
of X-ray binaries in globular clusters as a function of their Verbunt 
parameters, and (b) the evolution of compact-binary period distribution. 
In Sec.~\ref{cfobs}, we compare these model results with the current 
observational situation. Finally, we collect our conclusions and discuss 
future possibilities in Sec.~\ref{discuss}.  

\section{Model of Compact Binary Evolution in Globular Clusters}
\label{model}

We consider a binary population described by a number distribution $n(a,t)$, 
where $a$ is the binary separation, interacting with a fixed background of 
stars representing the core of a globular cluster of stellar density $\rho$ 
and core radius $r_c$. We now describe various ingredients of our model and 
the evolutionary scheme.

\subsection{Globular clusters} \label{gc}

Globular cluster cores are described by an average stellar density $\rho$, 
a velocity dispersion $v_c$, and a core radius $r_c$. In this work, we 
consider star-star and star-binary encounters of various kinds, but neglect 
binary-binary encounters. For characterizing the former two processes, two 
encounter rates are defined and used widely \citep{v2002,v2006}.
The first is the two-body stellar encounter rate $\Gamma$, 
which scales with $\rho^2r_c^3/v_c$, and occurs naturally in the rates of 
two-body processes like tidal capture, stellar collisions and merger. 
In fact, we can \emph{define} it as 
\begin{equation} 
\Gamma\equiv{\rho^2r_c^3\over v_c}\propto\rho^{3/2}r_c^2,
\label{eq:Gamma}
\end{equation}
for our purposes here. Note that the last scaling in the above equation holds
only for virialized cores, where the scaling $v_c\propto\rho^{1/2}r_c$ can be
applied. In this work, we shall use this assumption where necessary, but
with the caveat that some observed globular clusters have clearly not 
virialized yet.

The second is a measure of the rate of encounter between binaries and single 
stars in the cluster, the rate normally used being the encounter rate 
$\gamma$ of a \emph{single} binary with the stellar background, with the 
understanding that the total rate of binary-single star encounter in the 
cluster will be $\propto n\gamma$. We can \emph{define} $\gamma$ for our 
purposes as we did in \citet{bg2006}, namely,
\begin{equation} 
\gamma\equiv{\rho\over v_c}\propto\rho^{1/2}r_c^{-1}, 
\label{eq:gamma}
\end{equation}
where the last scaling holds, again, only for virialized cores. 

The importance of the above cluster parameters $\Gamma$ and $\gamma$ in this 
context has been extensively discussed by Verbunt \citep{v2002,v2006}, and 
we shall call them \emph{Verbunt parameters} here. Note that, for virialized 
cores, we can invert Eqs.~(\ref{eq:Gamma}) and (\ref{eq:gamma}) to obtain
the scaling of the core density and radius with the Verbunt parameters as:
\begin{equation} 
\rho\propto\Gamma^{2/5}\gamma^{4/5}, \qquad
r_c\propto\Gamma^{1/5}\gamma^{-3/5}
\label{eq:invert} 
\end{equation}

\begin{figure}
\epsscale{1.0}
\plotone{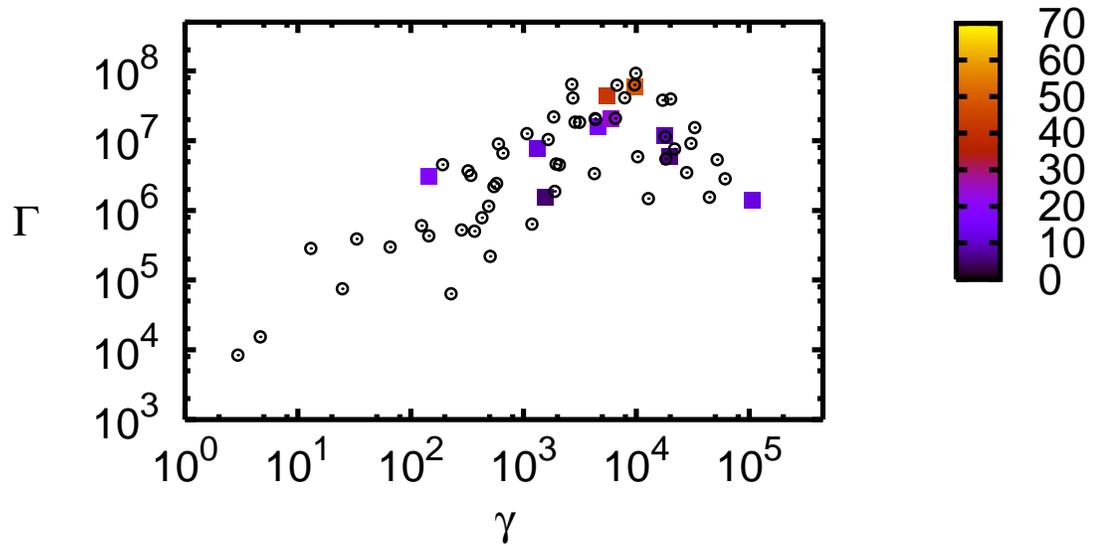}
\caption{Positions of Galactic globular clusters (open circles with
dots) on the $\Gamma-\gamma$ (Verbunt parameters) plane. Overplotted 
are positions of those clusters with significant numbers of X-ray 
sources detected in them (filled squares), color-coded according to 
the number of X-ray sources in each, the color code being displayed 
on the right. Data from \citet{h96}.}
\label{fig1}
\end{figure}

It is most instructive to display the observed globular clusters in the  
$\Gamma - \gamma$ plane, which we do\footnote{Alternatively, the display can 
be in the $\rho - r_c$ plane, as in Verbunt's original work. We find the 
cluster dynamics more transparent when shown directly in terms of the 
Verbunt parameters.} in Fig.~\ref{fig1}. The point that immediately strikes 
one in the figure is that the observed globular clusters seem to occur in a 
preferred, diagonal, ``allowed'' band in the $\Gamma - \gamma$ plane, along 
which there is a strong, positive correlation between the two parameters. 
We shall return to the significance of this elsewhere. 

In Fig.~\ref{fig1}, we also overplot the positions of those clusters in 
which significant numbers of X-ray sources have been detected, color-coding 
them according to the number of X-ray sources in each of them, as indicated. 
It is clear that these clusters are all in the upper parts of the above 
``allowed'' band, which is entirely consistent with the widely-accepted 
modern idea that the dominant mechanisms for forming these compact XBs 
in globular clusters are dynamical, \eg, tidal capture, exchange 
encounters, and so on, since such mechanisms occur more efficiently at 
higher values of the Verbunt parameters $\Gamma$ and $\gamma$, 
corresponding to higher stellar densities in the cluster core. Note that
the probability of destruction of binaries by dynamical processes 
also increases with increasing $\gamma$, as we shall see below, so that, 
at first sight, we might have expected the highest incidence of XBs 
in those clusters which have high $\Gamma$ and low $\gamma$. 
However, since $\Gamma$ and $\gamma$ are strongly correlated positively, 
as above, we cannot have arbitrarily high $\Gamma$ and low $\gamma$ for 
the same cluster. In reality, the highest number of XBs seem to occur, as 
Fig.~\ref{fig1} shows, in those clusters which have the highest values
of $\Gamma$ and high, but not the highest, values of $\gamma$. We return 
to this point later in the paper, where we present our theoretical 
expectations for the scaling of the number of binary X-ray sources with 
the Verbunt parameters $\Gamma$ and $\gamma$ on the basis of the 
evolutionary scheme explored here.

In modeling the globular cluster core as a static background in this work, 
we assume that, initially, a fraction $k_b$ of the stars is in primordial 
binaries, and that a fraction $k_X$ of the stellar population is compact, 
degenerate stars with the canonical mass $m_X=1.4\Ms$ (representing neutron 
stars and heavy white dwarfs). The rest of the stellar background 
(including the primordial binaries) is taken to consist of low-mass stars 
of the canonical mass $m_f=0.6\Ms$, which is a reasonable estimate of the 
mean stellar mass of a mass-seggregated core \citep{z197}. Naturally, the 
compact binaries formed from these ingredients consist of a degenerate 
star of mass $m_X=1.4\Ms$, and a low-mass companion of mass 
$m_c=m_f=0.6\Ms$. While this is clearly an oversimplification which must 
be improved upon in subsequent work, it appears to be adequate for a first 
look, which is our purpose here.

\subsection{A Boltzmann evolutionary scheme} \label{evol}
    
We explore in this work a Boltzmann evolutionary scheme, wherein the
evolution of the number $n(a,t)$ of binaries per unit interval in the 
binary separation $a$ (we choose to work here with $a$; equivalent 
descriptions in terms of the binary period $P$ or the internal binding 
energy [see Sec.~\ref{intro}] $E_{in}$ are possible, of course) is 
described by
\begin{equation}
\frac{Dn(a,t)}{Dt}=R(a) - nD(a).
\label{eq:Dn}
\end{equation}  
Here, $Dn(a,t)/Dt\equiv\partial n/\partial t + (\partial n/\partial a)
(da/dt)$ is the total derivative of bivariate $n(a,t)$: as explained in 
Sec.~\ref{intro}, this $n(a,t)$ is the result of an integration of a 
general, multivariate binary distribution over the variables we do not 
follow explicitly in this study, \eg, the external binding energy or, 
equivalently, the position of the binary inside the globular cluster. 
Further, $R(a)$ is the total rate of binary formation per unit interval 
in $a$ due to the various processes detailed below, and $D(a)$ is the 
total rate of binary destruction \emph{per binary} per unit interval in 
$a$ due to various processes, also detailed below. As our model stellar 
background representing the cluster core is taken as static for Papers I 
and II, the Verbunt parameters $\Gamma$ and $\gamma$ are time-independent, 
so that the formation and destruction rates $R$ and $D$ only depend on 
$a$ and the stellar masses. 

The above evolution equation can be re-written in the usual Boltzmann form
\begin{equation}
\frac{\partial n}{\partial t} = R(a) - nD(a) - \frac{\partial n}
{\partial a}f(a),
\label{eq:Evol}
\end{equation}
where $f(a)\equiv da/dt$ represents the total rate of shrinkage or 
\emph{hardening} of binaries (\ie, $da/dt<0$) due to several effects, which
we introduced in Sec.~\ref{intro}, and which we elaborate on below. In the
absence of all processes of formation and destruction, $R(a)=0=D(a)$, 
Eq.~(\ref{eq:Evol}) becomes the usual collisionless Boltzmann equation
\begin{equation}
\frac{\partial n}{\partial t} = - \frac{\partial n}{\partial a}f(a),
\label{eq:cless}
\end{equation}
representing a movement or ``current'' of binaries from larger to smaller
values of $a$ due to hardening. Equation~(\ref{eq:cless}) as akin to a wave 
equation with a formal ``phase velocity'' $f(a)$ of propagation. This 
analogy often proves useful for solving many problems, even with the 
more complicated formation and destruction terms present in 
Eq.~(\ref{eq:Evol}). Note that, when $f(a)$ is constant (or roughly so, 
which can happen under certain circumstances, as we shall see later), the 
elementary wave-equation analogy is quite exact, and solutions of the form
$n(a-f_0t)$ should apply. We shall explore this point elsewhere.

Note further that the Boltzmann scheme outlined above does not have an
explicit inclusion of the escape of those binaries from the globular cluster  
which receive a sufficiently large ``kick''. In principle, we can include
this by suitably generalizing the above destruction term $D(a)$. However,
in this introductory study, this did not appear crucial, as the main 
population affected by this process is that of primordial binaries, whereas
our main concern here is with dynamically-formed compact binaries. The 
latter are, generally speaking, already so hard at formation that this 
process is much less effective in ejecting them from the cluster. 
Accordingly, we neglect this process here.    

\subsection{Binary hardening processes} \label{hard}  

In all of the dynamical encounter processes considered in this work, \viz, 
collisional hardening (described in this subsection), and dynamical formation
and destruction processes (described in the next subsections), we shall assume
the orbits to be circular, \ie, neglect their eccentricity. This is, again, a 
simplification used for a first look. However, it is well-known from extensive
numerical simulations that a large majority of the binaries formed by tidal 
capture are circular or nearly so \citep{z297}, due to the rapid 
circularization which follows capture. Since our main concern here is with 
dynamically-formed binaries, this approximation may well be a reasonable one 
for describing overall evolutionary properties of such binary populations.
   
\subsubsection{Hardening in pre-X-ray binary (PXB) phase}\label{pxbhard}  

As explained in detail in  \citet{bg2006}, referred to henceforth as BG06,
the processes that harden binaries are of two types, \viz, (a) those which
operate in isolated binaries, and are therefore always operational, and
(b) those which operate only when the binary in a globular cluster. In the
former category are the processes of gravitational radiation and magnetic
braking, and in the latter category is that of collisional hardening. As 
discussed in detail in BG06, collisional hardening, which increases with
increasing $a$, dominates at larger orbital radii, while gravitational 
radiation and magnetic braking, which increase steeply with decreasing $a$, 
dominate at smaller orbital radii.  It is these processes that harden a 
compact binary from its pre-X-ray binary (PXB) phase, during which its
orbit is still not narrow enough for the companion (mass donor) star to
come into Roche lobe contact, to the state where this Roche lobe contact
does occur, at which point the companion starts transferring mass to the  
degenerate star, and the system turns on as an X-ray binary (XB) ---
either a CV or a LMXB, depending on the nature of the degenerate accretor. 

Consider gravitational radiation first. The relative angular momentum loss 
rate due to this process is:
\begin{equation}
j_{GW}(a) \equiv
\left( \frac{\dot J}{J} \right)_{GW} = -\alpha_{GW}a^{-4}, \qquad
\alpha_{GW} \equiv \frac{32G^3}{5c^5} m_c m_X (m_c + m_X) 
\label{eq:GW}.
\end{equation}
Here, as before, $m_X$ is the mass in solar units of the degenerate primary 
(neutron star or white dwarf) which emits X-rays when accretion on it occurs 
during the mass-transfer phase of the compact binary, $m_c$ is the mass 
of its low-mass companion in solar units, and the unit of the binary orbital
radius $a$ is the solar radius. We shall use these units throughout the work.

Now consider magnetic braking. The pioneering Verbunt-Zwaan 
\citep{vz81} prescription for this process has been reassessed  and
partly revised in recent years, in view of further observational evidence
on short-period binaries available now (for further details, see discussions 
in BG06 and references therein), and modern prescriptions are 
suggested in \citet{slu1}. From these, we have chosen for this work the
following one which preserves the original Verbunt-Zwaan scaling, but 
advocates an overall reduction in the strength of the magnetic braking
process:    
\begin{equation}
j_{MB}(a) \equiv
\left( \frac{\dot J}{J} \right)_{MB} = -\alpha_{MB}a^{-5}, \quad
\alpha_{MB} \equiv 9.5\times10^{-31}G R_c^4 \frac{M^3}{m_X m_c}, 
\quad M \equiv m_c + m_X
\label{eq:MB}
\end{equation}
Here, $R_c$ is the radius of the companion. 
Note that the strength of magnetic braking is still a matter of some 
controversy; while the evidence cited in the above reference argues for
a reduction from the original value, it can also be argued that the 
presence of the well-known ``period gap'' in the period distribution of
CVs requires a strength comparable to the original one. We have adopted    
here a recent prescription which is reasonably simple and adequate for 
our purposes: our final results do not depend significantly on the 
strength of this process.

Consider finally collisional hardening. As indicated earlier, it is a 
stochastic process, for whose continuous limit we use the prescription of
\citet{sh79}, as has been done previously in the literature (see BG06 for
a discussion). According to this prescription, the rate of increase of 
orbital binding energy $E$ of a compact binary due to collisional 
hardening is given in this limit by:
\begin{equation}
\left( \frac{\dot E}{E} \right)_{C} = A_{C} a \gamma , \qquad
A_{C} \equiv 18G \frac{m_f^3}{m_c m_X}  
\label{eq:COLLe}
\end{equation} 
Here, $m_f$ is the mass of the stars in the static background representing
the cluster. We shall use $\Ms{\rm pc}^{-3}$ and km sec$^{-1}$ as the 
units of $\rho$ and $v_c$ respectively. In the above units, the value of 
$\gamma$ for Galactic globular clusters typically lie between $\sim 10^3$ 
and $\sim 10^6$ (BG06). The relation between $\dot E$ and $\dot J$ is:
\begin{equation} 
\frac{\dot J}{J} = -\frac{1}{2}\frac{\dot E}{E} 
+ \frac{3}{2}\left( \frac{\dot m_c}{m_c} + \frac{\dot m_X}{m_X} \right),
\label{eq:EJ}
\end{equation}
and the angular momentum loss rate is related to the shrinkage rate of the 
orbit $\dot a$, or hardening, as:
\begin{equation}
\frac{\dot a}{a} = 2\frac{\dot J}{J} -2\frac{\dot m_c}{m_c} 
-2\frac{\dot m_X}{m_X}
\label{eq:Ja}
\end{equation}
The $\dot m_c$ and $\dot m_X$ terms on the right-hand side of 
Eqn.~(\ref{eq:Ja}) are nonzero during mass transfer in the XB phase. In the 
PXB phase, $\dot m_c=\dot m_X = 0$, so that $\dot a$ is simply related to 
$\dot J$ as (see BG06 and references therein):
\begin{equation}
\frac{\dot a}{a} = 2\frac{\dot J}{J}
\label{eq:adot_PXB}
\end{equation}
Using Eqns.~(\ref{eq:EJ}) and (\ref{eq:COLLe}), we have in this case,
\begin{equation}
j_{C}(a) \equiv
\left( \frac{\dot J}{J} \right)_C= -\frac{1}{2}\left( \frac{\dot E}{E} 
\right)_C
= \alpha_{C} a \gamma , \quad \alpha_{C} \equiv \frac{A_C}{2} =
9G \frac{m_f^3}{m_c m_X}  
\label{eq:COLL}
\end{equation}

The total rate of loss of orbital angular momentum due to the above
three processes is:
\begin{equation}
j_{TOT}(a) \equiv
\left( \frac{\dot J}{J} \right)_{TOT} = j_{GW}(a) + j_{MB}(a) + j_{C}(a)
\label{eq:TOT}
\end{equation}

\subsubsection{Hardening in X-ray binary (XB) phase}\label{xbhard}  

As mass transfer starts upon Roche lobe contact, its effect on the 
angular momentum balance in the XB must be taken into account, in the 
manner described below. Note first that, for the radius of the Roche-lobe 
$R_L$ of the companion, we can use either the 1971 Paczy\'nski  
approximation:
\begin{equation}
R_L/a = 0.462\left({m_c\over M}\right)^{1/3},
\label{eq:rochepaczy}
\end{equation}
which holds for $0<m_c/m_X<0.8$, or the 1983 Eggleton approximation:    
\begin{equation}
R_L/a = {0.49\over 0.6+q^{2/3}\ln(1+q^{-1/3})},
\qquad q\equiv m_X/m_c,  
\label{eq:rocheggle}
\end{equation}
which holds for the entire range of values of the mass ratio $q$. Both
approximations have been widely used in the literature, and they give
essentially identical results for the mass ratios of interest here.
We have used the Paczy\'nski approximation here for simplicity of
calculation.

At the Roche-lobe contact point, $R_L$ must be equal to the companion
radius, the value of which is $R_c\approx0.6\Rs$ for a companion of 
$m_c=0.6\Ms$ (see above), according to the mass-radius relation for low 
mass stars \citep{g2007}. 
For $m_X=1.4\Ms$, this translates into an orbital radius of
$a_L=1.94\Rs$ at Roche lobe contact, using Eqn.~(\ref{eq:rochepaczy}). 
After this, the companion continues to remain in Roche-lobe contact as 
the binary shrinks further, and continues to transfer mass 
\citep{heu91,heu92}. In other words, we have
\begin{equation}
R_c=0.46a\left({m_c\over M}\right)^{1/3}, \qquad (a<a_L)
\label{eq:Roche}
\end{equation}
throughout the XB phase. During this phase, the binary is already 
narrow enough that the collisional hardening rate is quite negligible 
compared to those due to gravitational radiation and magnetic braking. 

Since no significant mass loss is expected from the XB in this phase, 
we have  
\begin{equation}
\dot m_c = -\dot m_X.
\label{eq:mdot}
\end{equation}
Combining Eqns.~(\ref{eq:Ja}), (\ref{eq:Roche}) and (\ref{eq:mdot}) 
with a mass-radius relation for the companion of the form
\begin{equation}
R_c \propto m_c^s,
\end{equation}
we find:
\begin{equation}
\dot a=\frac{j_{tot}(a) a \left(s-\frac{1}{3}\right)}
{\left[\frac{s}{2} + \frac{5}{6} -\left(\frac{m_c}{M-m_c}\right)\right]}
\end{equation}
Here,  $j_{tot}(a) = j_{GW}(a) + j_{MB}(a)$ is the effective total rate 
of loss of angular momentum, since the collisional-hardening contributions 
are negligible, as explained above.

\begin{figure}
\plotone{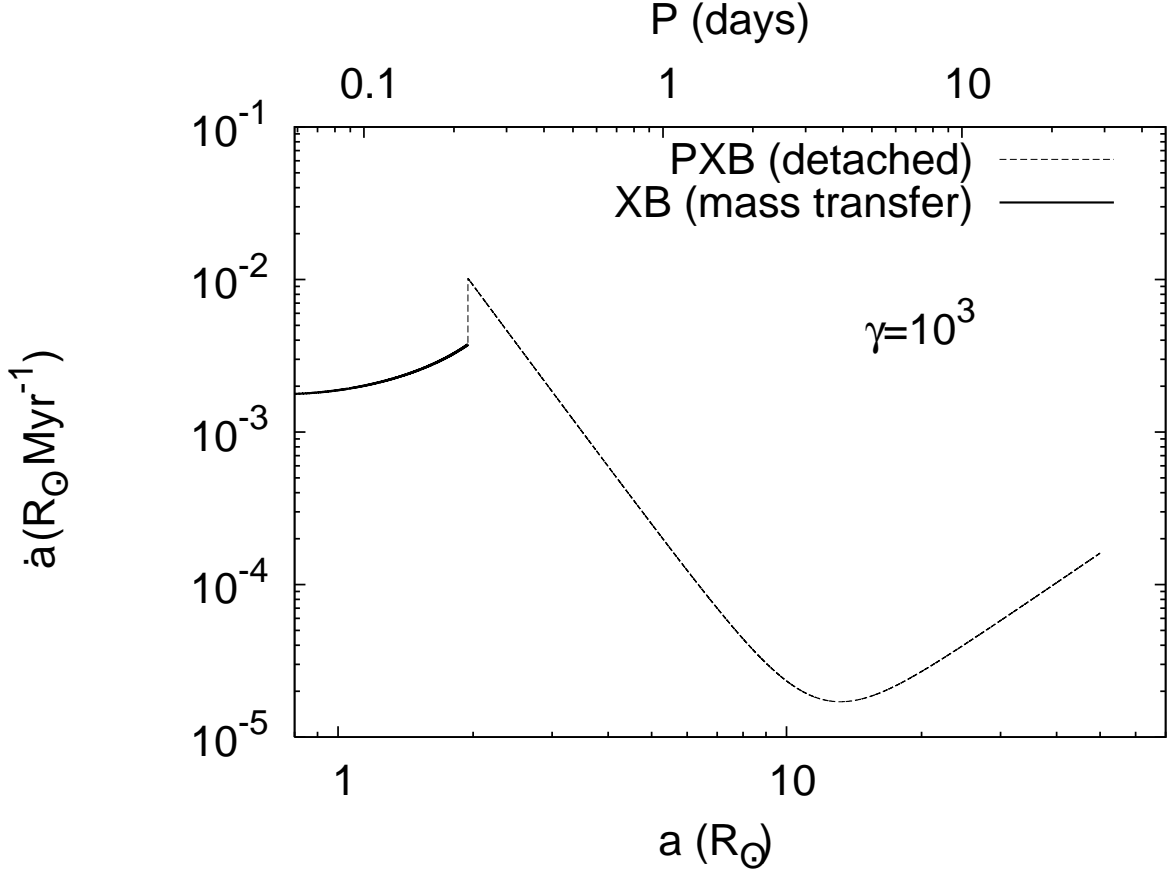}
\caption{Hardening rate $\dot a$ of a compact binary as a function of the 
orbital radius $a$, in a globular cluster with a Verbunt parameter of 
$\gamma=10^3$. Collisional hardening dominates roughly at $a>14\Rs$,
and gravitational radiation plus magnetic braking roughly in the range 
$2\Rs<a<14\Rs$. These two regions, shown as dashed lines, are populated
by pre-X-ray binaries (PXBs), which are detached. At $a\approx 2\Rs$, 
Roche lobe contact occurs and mass transfer begins, so that the region
shortward of this radius, shown as the solid line, is populated by X-ray 
binaries (XBs). This region is shown upto the orbital radius $a_{pm}$ 
which corresponds to the period minimum of $\approx 80$ min (see text).
Along abscissa, both orbital radius $a$ and orbital period $P$ scales 
are shown for convenience.}
\label{fig:adot}
\end{figure}

For the low-mass main sequence companions that we consider here, 
$s\approx 1$. However, when the mass of the companion becomes less than 
about 0.03\Ms, it becomes degenerate, so that $s\approx -1/3$ 
\citep{g2007}. This results in a widening of the orbit ($\dot a > 0$)
from this point onwards, which we do not follow here, since our study
is not aimed at such systems, as explained in Sec.~\ref{apply}. This 
change-over point is, of course, that corresponding to the well-known 
period minimum of $\approx 80$ minutes in the orbital evolution of CVs 
and LMXBs\citep{heu92}. Henceforth, we denote the value of $a$ at the
period minimum by $a_{pm}$, and we terminate the distributions of 
$\dot a$ and $n(a,t)$ in $a$ at a minimum value of $a_{pm}$ in the figures
shown in this work. Thus, in Fig.~\ref{fig:adot}, we display the 
hardening rate $\dot a$ against $a$, beginning from a wide PXB phase, 
going into Roche lobe contact, and continuing through the mass-transfer 
XB phase upto the above period minimum. Note that $\dot a$ has a very 
weak dependence on $a$ during the XB phase, which may have interesting 
consequences, as we shall see later.
 
\subsection{Binary formation processes}\label{form}

Compact binaries with degenerate primaries and low-mass companions are 
formed in globular cluster (henceforth GC) 
cores primarily by means of two dynamical 
processes, namely, (i) tidal capture (tc) of a degenerate, compact star 
(white dwarf or neutron star) by an ordinary star, and (ii) an exchange 
encounter (ex1) between such a compact star and a binary of two ordinary 
stars, wherein the compact star replaces one of the binary members. 
Accordingly, the total rate of formation of compact binaries per unit 
binary radius, $R(a)$, consists of the above tc rate $r_{tc}(a)$ and ex1 
rate $r_{ex1}(a)$:
\begin{equation} 
R(a)=r_{tc}(a) + r_{ex1}(a)
\label{eq:r_tot}
\end{equation}
where $a$ is the orbital radius of the compact binary so formed. We now 
consider the rates of formation by tidal capture and by exchange.  
 
\subsubsection{Tidal capture}\label{tcform} 

In a close encounter between a compact star of mass $m_X$ and an ordinary 
star of mass $m_c$ with a distance of closest approach $r_p$, tidal 
capture can occur if their relative speed $v$ is less than an appropriate 
critical speed $v_0(r_p)$, which we discuss below. The cross section for 
encounters within this distance $r_p$ is given by the well-known form 
\citep{spz}:
\begin{equation}
\sigma_g=\left( \pi r_p^2 + \frac{2\pi GMr_p}{v^2} \right)
\label{eq:gf}
\end{equation} 
which gives the \emph{differential} cross section for tidal capture around 
$r_p$ as:  
\begin{equation}
{d\sigma_{tc}\over dr_p} = \left\lbrace 
\begin{array}{ll}
\left( 2\pi r_p + \frac{2\pi GM}{v^2} \right)dr_p & v<v_0(r_p)\\
0	& v \geq v_0(r_p) 
\end{array}
\right .
\label{eq:s_tc1}
\end{equation}
The first terms in the right-hand sides of Eqs.~(\ref{eq:gf}) and 
(\ref{eq:s_tc1}) are the obvious geometrical cross sections and the 
second terms are due to gravitational focusing (also see below). It is clear 
that the latter terms dominate when $r_p$ is small, as is the case for the
range of values of $r_p$ relevant to the problem we study here. We shall
return later to the actual numerical values of $r_p$ of interest to us in this
study.

After being tidally formed, the binary is believed to circularize very 
rapidly to an orbital radius $a=2r_p$, assuming conservation of angular 
momentum \citep{spz}. Accordingly, the differential cross-section in 
terms of $a$ is given by:
\begin{equation}
{d\sigma_{tc}\over da} = \left \lbrace 
\begin{array}{ll}
\left( \frac{\pi}{2}a + \frac{\pi GM}{v^2} \right) & v<v_0(a)\\
0	& v \geq v_0(a) 
\end{array}
\right .
\label{eq:s_tc2}
\end{equation}
Here,  $v_0(a)$ is the critical velocity in terms of $a$, obtained by setting
$r_p=a/2$ in Eq.~(\ref{eq:v01}) below.

In a sense, the whole cross-section as expressed above may be regarded
as ``geometrical'', if we look upon pure considerations of 
Newtonian gravity as being geometrical. Details of the essential 
astrophysics enter only when we calculate the critical speed $v_0(r_p)$,
and an inversion of this relation (together with other plausible 
requirements; see below) then readily gives us the range of $r_p$ over 
which tidal capture is physically admissible. This is an interesting
topic, with literature going back to the mid-1970s and earlier, and we 
summarize in this section those essential points which we need in this 
work. The basic physics of tidal capture is of course that, during a close 
encounter, the degenerate compact star excites non-radial oscillation  
modes in the normal companion star through tidal forcing (in an 
encounter between two normal stars, each excites oscillations in the 
other): the energy required to excite these oscillations comes from
the kinetic energy of relative motion of the two stars, so that if
enough energy is extracted from this source by exciting these modes,
the stars become bound after the encounter. This energy condition 
readily translates into one between $v_0$ and $r_p$, giving an upper
limit $v_0$ on velocity for a specified $r_p$ as above, or, as  
expressed more commonly, an upper limit on the distance of closest
approach $r_p$ for a specified velocity (actually, often a 
\emph{distribution} of velocities, \eg, a Maxwellian, with a specified 
parameter in practical situations, as we shall see below).      

The above relation between $v_0$ and $r_p$ has been calculated in 
the literature at various levels of detail. The pioneering estimates 
given in \citet{fpr75} or earlier works basically employ the impulse 
approximation for calculating the gain in the internal energy of the 
tidally-perturbed star, wherein the changes in the positions of the 
two stars during the tidal interaction are neglected. A clear account 
of the procedure is given in \citet{spz}, where the final result is
evaluated for two normal stars of equal masses. Upon generalizing this
procedure appropriately to the problem we study, where we have (a) 
unequal stellar masses $m_X$ and $m_c$, and (b) the fact that only 
the normal star of mass $m_c$ undergoes tidally-induced oscillations,
we obtain the following relation between $v_0$ and $r_p$:
\begin{equation}
v_0(r_p)=\left( \frac{4}{3}G m_X R_m^2 \right)^
\frac{1}{2}r_p^{-\frac{3}{2}}
\label{eq:v01}
\end{equation} 
Here, $R_m$ is the root-mean-square radius of the companion star, \ie, 
its radius of gyration which is given in the polyrtopic approximation 
as $R_m^2/R_c^2\approx 0.114$ in terms of the companion's radius 
$R_c$ \citep{spz}.  

To obtain the overall rate of tidal capture in the GC core of volume  
$4\pi r_c^3/3$ per unit interval in $a$ around $a$, we first consider 
this rate around a particular value $v$ of the above relative velocity of 
encounter, \ie, $r_{tc}(a,v) = (4\pi/3)r_c^3 k_X \rho^2(d\sigma_{tc}/da)v$, 
in terms of the above differential cross-section, remembering that the rate 
of encounter scales with the product of the densities $k_X \rho$ and 
$\rho$ of compact stars and normal stars respectively. We then average 
this rate over the distribution of $v$, obtaining the form:
\begin{equation}
r_{tc}(a)=\frac{4}{3}\pi r_c^3 k_X \rho^2 \langle\sigma_{tc}(a,v)v\rangle,
\label{eq:r_tc}
\end{equation}  
where the angular brackets indicate an average over the $v$-distribution.

For the actual averaging, we adopt in this work a Maxwellian distribution 
$f_{mx}(v)$, as has been widely done in the literature. A normalized 
Maxwellian is
\begin{equation}
f_{mx}(v) = Av^2\exp(-\beta v^2),\quad \beta\equiv\frac{3}{2v_c^2}, \quad
A\equiv\frac{4}{\sqrt{\pi}}\beta^{3\over2},
\label{eq:max}
\end{equation}
where $v_c$ is the velocity dispersion introduced earlier, for which we 
adopt the canonical value 10 km s$^{-1}$ in the numerical calculations (also
see below).  

With the aid of Eqns.~(\ref{eq:s_tc2}), (\ref{eq:v01}) and (\ref{eq:max}), 
we perform the averaging and obtain:
\begin{equation}
\begin{array}{l}
\langle\sigma_{tc}(a,v)v\rangle=I_{geo}+I_{grav},\\
{\rm ~ where},\\
I_{geo}\equiv\sqrt{\pi\over\beta}a\left[1-\exp(-\beta v_0^2(a))
(\beta v_0^2(a)+1)\right]\\
I_{grav}\equiv2\sqrt{\pi}GM\beta^{1\over2}
\left[1-\exp(-\beta v_0^2(a))\right]
\end{array}
\label{eq:sv}
\end{equation}
The terms $I_{geo}$ and $I_{grav}$ above arise due to what we described
respectively as the geometrical term and the gravitational focusing term in
the discussion below Eq.~(\ref{eq:s_tc1}). Eqns.~(\ref{eq:r_tc}) and 
(\ref{eq:sv}) together give the total tidal capture rate as:
\begin{equation}
r_{tc}(a)=\sqrt{\frac{32\pi^3}{3}}k_X\Gamma GM
\left[1-\exp(-\beta v_0^2(a))\right],
\label{eq:r_tc2}
\end{equation}
where $\Gamma$ is the Verbunt parameter describing the total two-body 
encounter rate in the cluster core, as introduced earlier, and we have 
ignored $I_{geo}$ compared to $I_{grav}$, which is an excellent 
approximation for the range of $r_p$ or $a$ relevant here.

\begin{figure}
\plotone{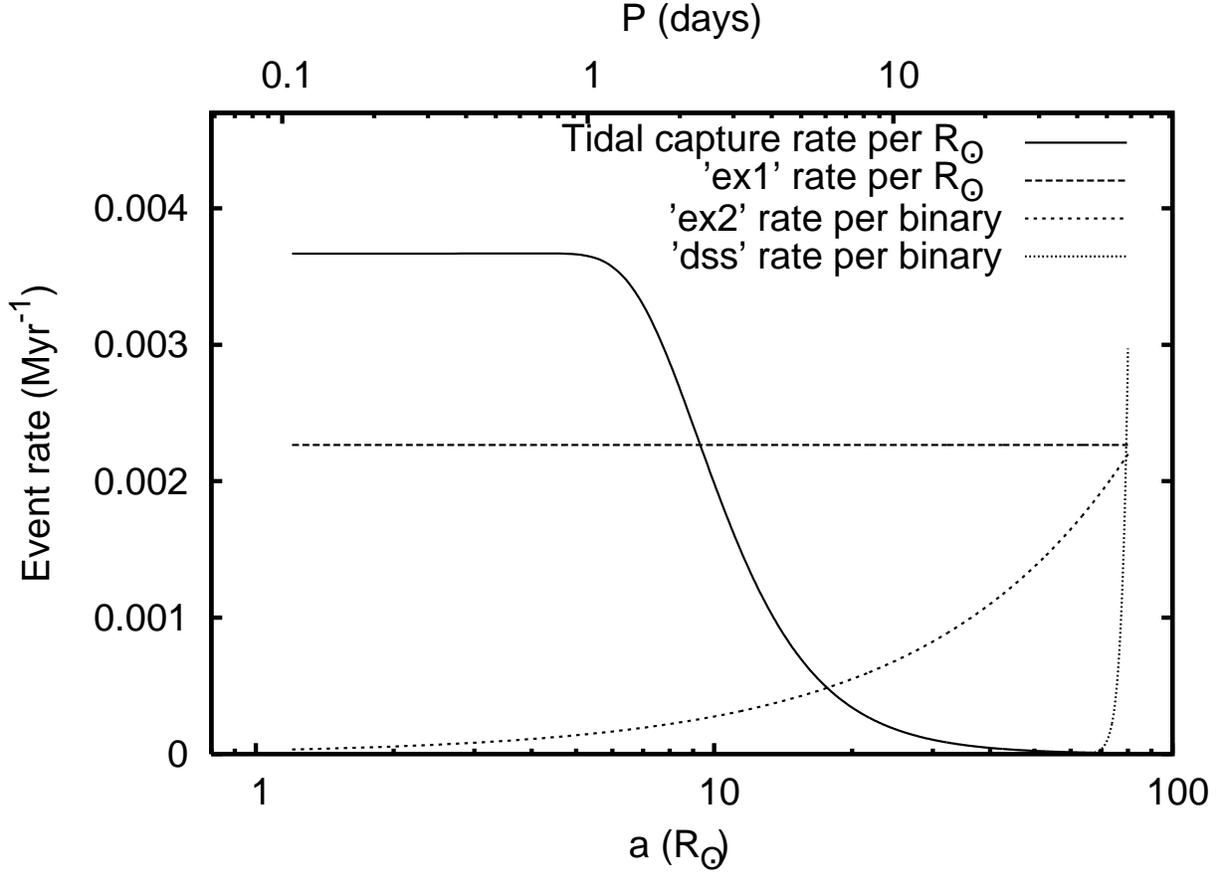}
\caption{Tidal capture (tc) rate, the exchange rates `ex1' and `ex2',
and the dissociation (dss) rate, as described in text. Note that, 
compared to the tc rate, the ex1 rate has been magnified by a factor
of 50, the ex2 rate rate by a of factor 60, and the dss rate by a 
factor of $10^9$, so that all rates are clearly visible. Along 
abscissa, both orbital radius $a$ and orbital period $P$ scales are 
shown for convenience. Curves are terminated at a radius $a_{min}
= 1.2\Rs$ (see text).}
\label{fig:rates}
\end{figure}

We show in Fig.~\ref{fig:rates} $r_{tc}$ given by Eq.~(\ref{eq:r_tc2})
as a function of $a$: this tidal capture cross-section is nearly 
constant for $a<5\Rs$, and decreases rapidly at larger $a$. At this 
point, we need to invoke additional physical arguments in order to
estimate the range of values of $a$ or $r_p$ over which tidal capture is 
actually possible, and use the above cross-section only over this range
for our calculations. The lower bound to the above range comes from the 
requirement that the two stars must form a binary and not merge into each 
other, and the upper bound comes from the requirement introduced earlier   
that enough energy of relative motion between the two stars must be  
absorbed by the tidally-excited oscillation modes that the stars become
bound. Consider the lower bound on $r_p$ first. Clearly, a minimum value
of this bound must be the sum of the stellar radii, which in our case leads
to the bound $r_p\ge R_c\approx 0.6\Rs$. A more conservative bound 
comes from the requirement that the companion must underfill its Roche 
lobe after the binary has formed, \ie, $R_c\le R_L$, which, with the aid of 
Eq.~(\ref{eq:rochepaczy}) and $a=2r_p$, yields $r_p\ge 1.6R_c\approx
\Rs$ for the masses $m_X=1.4\Ms$ and $m_c=0.6\Ms$ we have here. 
The idea behind the latter requirement is apparently that if the companion
overfills its Roche lobe at this point, the ensuing mass transfer is likely to
lead to a merger. This seems reasonable at first, but detailed N-body 
simulations of recent years have suggested that this requirement may, in
fact, be too restrictive. In the simulations of  \citet{z297}, which included
stellar evolutionary effects according to the scheme of these authors, 
systems which violated the latter requirement but satisfied the former one 
were allowed to evolve, with the result that details of the evolution 
determined which systems merged and which did not. In fact, these 
authors found a lower limit on $a=2r_p$ of approximately $a\ge \Rs$ for 
tidal capture with an average companion mass very similar to ours, which 
is to be compared with the limits $a\ge 1.2\Rs$ from the first requirement 
above, and $a\ge 2\Rs$ from the latter. In view of this, we have adopted
the lower bound of $a^{min}\approx1.2\Rs$ for our calculations here, as 
shown in Fig.~\ref{fig:rates}.

Consider now the upper bound on $r_p$. We have already given the 
relation between $r_p$ and $v_0$ by Eq.~(\ref{eq:v01}) in the impulse
approximation. Remembering that $v_0^2 = 1/\beta = 2v_c^2/3$ for a
Maxwellian, the above relation yields, for a canonical value 
$v_c$ = 10 km s$^{-1}$ as given above, an upper limit of
$r_p\le 10.2R_c$ for a polytropic index $n=3$ and one of  
$r_p\le 14.1R_c$ for $n=1.5$. Note that these bounds of $r_p/R_c$ are 
larger than those given for two stars of equal mass (roughly 8 for 
$n=3$ and 11 for $n=1.5$) in Table 6.2 of \citet{spz} by a factor of 
$(m_X/m_c)^{1/3}$ since $r_p^{max}/R_c$ scales with the mass-ratio in 
this manner in the impulse approximation, as can be seen readily from 
Eq.~(\ref{eq:v01}), remembering that $R_c\propto m_c$ for the companions 
we consider here. That $r_p^{max}/R_c$ should increase with increasing 
$(m_X/m_c)$ is qualitatively quite obvious, since, other things being 
equal, a higher value of the mass ratio excites tidally-forced 
oscillations of larger amplitude. We return below to the question of the 
exact scaling with this mass ratio.  

As has been realized long ago, the impulse approximation is of limited
validity, working best when the frequency of perturbation (\ie, tidal 
forcing) is not very different from those of the stellar oscillation modes 
that are excited by this perturbation \citep{fpr75, spz}. Since this is not
the case for the values of $r_p^{max}/R_c$ estimated above, we need more 
accurate results, which come from detailed computations of the total 
energy dissipated by the above excited modes. Such numerical computations 
were pioneered by \citet{pt77}, and detailed results were established 
for various situations by several groups of authors in the mid-1980s,
including \citet{lo86} and \citet{mmt87}, which have been extensively
used since. These results have shown that the exact upper bounds on 
$r_p$ are considerably smaller than those given by the impulse 
approximation, as may have been expected, since the forcing frequency  
falls far below those of the oscillation modes at such large separations
as are given by this approximation, and the efficiency of exciting these 
modes drops rapidly. Some exact results are given in  Table 6.2 of 
\citet{spz} from the above references, but only for the equal-mass case,
where the above upper bound $r_p^{max}/R_c$ is 2.4 for $n=3$ and 3.4 for 
$n=1.5$. 

For our purposes here, we need to obtain the above upper bounds for our 
mass ratio $m_X/m_c=1.4/0.6\approx 2.3$, which we do by doing a power-law 
fit of the form $r_p^{max}\propto(m_X/m_c)^{\alpha}$ to the results given 
for various values of the degenerate/normal star mass-ratios in Table 3 
of \citet{lo86}. This yields $\alpha\approx 0.62$ (note that the quantity
listed in Table 3 of \citet{lo86} is the impact parameter $R_0$ defined      
by these authors; $r_p^{max}$ scales as $R_0^2$, as shown in their 
paper). The interesting point about this scaling is that it is stronger
than that given above by the impulse approximation, which corresponds to
$\alpha=1/3$. Clearly, then, the impulse approximation fails to extract
the entire scaling with $m_X/m_c$. The reason for this appears to be 
related to nonlinear effects in exciting and dissipating tidally-induced
oscillations, but needs to be investigated further\footnote{Note that 
this discrepancy is even stronger for the case where both stars are 
normal, main-sequence ones, since $\alpha\approx 1.6$ in that case, as
can be shown readily from Table 2 in the above Lee-Ostriker reference.  
An obvious line of reasoning for this would be that larger nonlinear
effects may be expected when two normal stars force tidal oscillations
in each other, but we shall not speculate on this any further here.}.
With the above value of $\alpha$, the upper bound $r_p^{max}/R_c$ for 
our mass-ratio here is 4.1 for $n=3$ and 5.7 for $n=1.5$. As the latter
value of the polytropic index is believed to give a better 
representation of a low-mass main-sequence companion of the kind we
are considering here, we adopt $r_p^{max}/R_c\approx 5.7$ here. With
$a=2r_p$ and the value of $R_c$ given earlier, this translates into an
upper bound on $a$ as $a^{max}\approx 6.8\Rs$, which we can adopt for 
these calculations.     

Thus we find a range of values $1.2\Rs\le a\le 6.8\Rs$ over which 
tidal capture is expected to be effective in the problem we study
here. Consider now how the tidal-capture cross-section is expected to
fall off at the bounds of this range. At the upper bound, the cut-off
is not sharp, of course, as there is a distribution of velocities. 
In other words, the upper bound $a^{max}$ as given above corresponds  
to a suitable average (actually, root-mean-square in this case) 
velocity, so that at any $a>a^{max}$, there will be some stars in the 
distribution whose velocities are sufficiently below this average 
that tidal capture will be possible for them. Of course, their number
will decrease as $a$ increases, producing a ``tail'' in the tidal 
capture cross-section whose shape is determined by that of the 
velocity distribution. We have used a Maxwellian distribution here, 
which gives the tail seen in Fig.~\ref{fig:rates}, which falls off 
rapidly beyond $a^{max}=6.8\Rs$. We shall use this fall-off profile in
our calculations: other profiles will not make a large difference.
At the lower bound, in view of the discussion given earlier, we expect 
the cross-section to actually fall off gradually from about $a=2\Rs$ 
to $a=a^{min}=1.2\Rs$, rather than being cut off sharply at $a^{min}$,   
but we shall ignore this complication here. 

We close this discussion of tidal capture with some observations on 
the many investigations, conclusions, and points of view that the 
subject has now seen for more than three decades. From the pioneering
suggestion and an essentially dimensional estimate of \citet{fpr75},  
detailed calculations of the 1980s and '90s have reached interesting,
and sometimes contradictory, conclusions. For example, concerns that 
energy dissipation by tidally-induced modes may lead to a large
distention of the companion and so to a merger have been confronted
with results from detailed computations of the nonlinear damping of
the primary modes by coupling to other, high-degree modes, which
suggested that the damping took place far more rapidly than thought
before, and the energy dissipated was too small to have a significant
effect on the companion's structure. We here have a adopted a 
somewhat moderate view that tidal capture is plausible, but efficient 
over only a restricted range of $r_p$ or $a$. This view is supported
by (a) recent observational demonstration that the number of X-ray
sources in Galactic globular clusters scale with their Verbunt 
parameter $\Gamma$, \ie, the two-body encounter rate \citep{pool03},
as described earlier, and (b) recent N-body simulations of 
\citet{z297} showing tidal capture over a considerable range of $a$,
admittedly under the algorithms adopted by these authors. Consider,
finally, our suggested range of radii for efficient tidal capture,
$a^{max}/a^{min} \approx 5.7$, 
as given above, in the context of 
other suggested ranges. Values in the range $a^{max}/a^{min}\approx 
2-3$ have been thought plausible by \citet{prp02}, while \citet{z297}  
have demonstrated tidal capture over a range $a^{max}/a^{min}\approx
10$. We here advocate a range $a^{max}/a^{min}\approx 4-6$ (depending
on $n$), which is between the two, and still quite modest. 

\subsubsection{Formation by exchange}\label{exform} 

Exchange encounters between binaries and single stars with arbitrary 
mass ratios has been extensively studied by \citet{hhm96}. They 
performed detailed numerical scattering experiments, using the 
automatic scattering tools of the STARLAB package. From the resulting 
exchange cross sections, they obtained a semi-analytic fit of the form:
\begin{equation}
\sigma_{ex}(R)=\frac{\pi GM_{tot}R}{2v^2}\overline\sigma(m_1, m_2, m_3).
\label{eq:s_ex}
\end{equation}
Here, $R$ is the orbital radius of the initial binary, $m_1$ is the 
mass of the escaping star, $m_2$ is the companion mass, $m_3$ is the 
mass of the incoming star, and $M_{tot}\equiv m_1+m_2+m_3$. $\sigma(m_1, 
m_2, m_3)$ is the dimensionless cross section which is a function of 
these masses only and which is given by Eq.~(17) of \citet{hhm96}. We 
use Eqn.~(\ref{eq:s_ex}) to obtain the cross sections 
$\overline\sigma_{ex1}(a)$ for the exchange process `ex1' described 
above, but one essential point needs to be clarified first.

\begin{figure}
\plotone{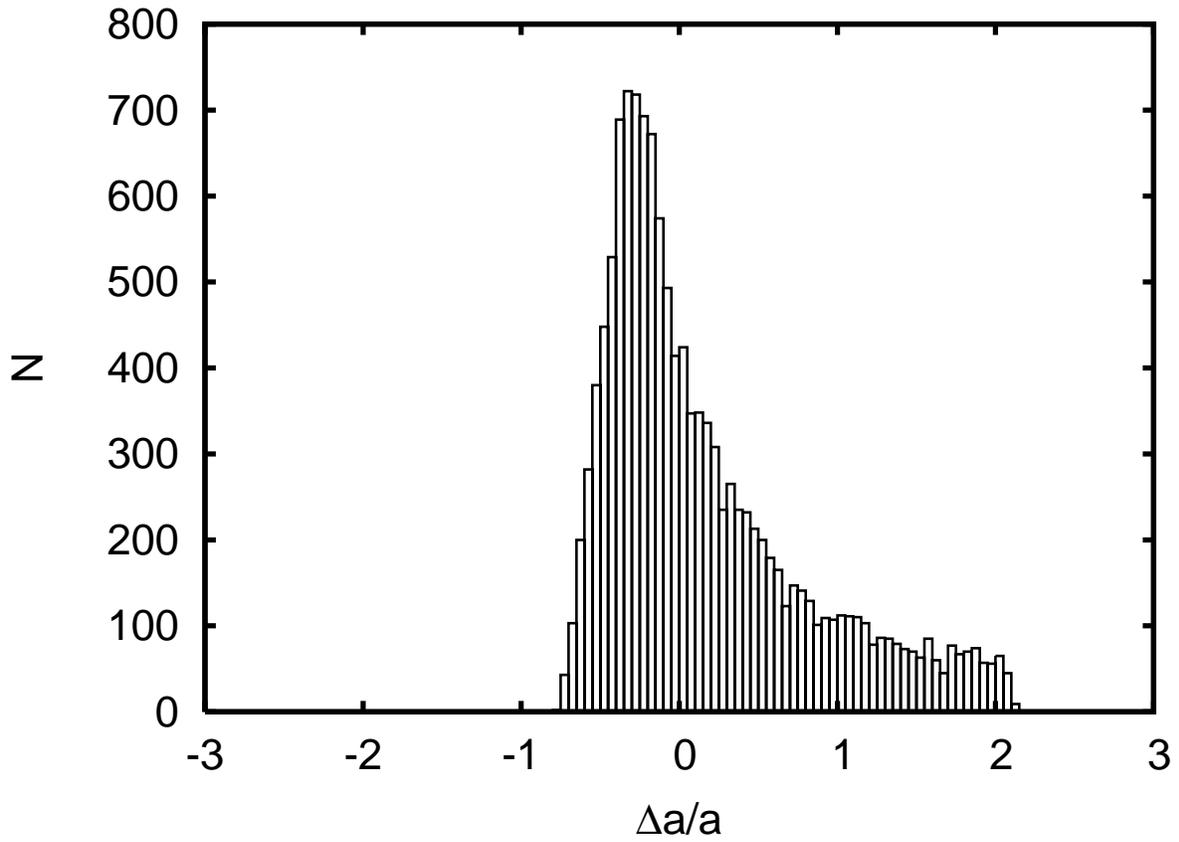}
\caption{Distribution of the fractional change in binary radius 
$\Delta a/a$ for $\sim 30000$ scattering experiments with 
$v/v_{crit}=0.5$ (see text) and random impact parameters. The 
distribution is highly asymmetric, with  a peak at $\Delta a/a
\approx -0.25$, and a long tail in the $\Delta a>0$ direction.}
\label{fig:delta}
\end{figure}

The radius $a$ of the compact binary formed by exchange is not the 
same as the radius $a^\prime$ of the original binary undergoing exchange. 
Therefore, a relation between $a^\prime$ and $a$ is required, since in 
Eqn.~(\ref{eq:s_ex}) $R$ represents the radius $a^\prime$ of the initial 
binary, \emph{not} the radius $a$ of the compact binary formed by 
exchange. According to the binary-hardening rule of Heggie \citep{h75}, 
the final compact binary must, on an average, be harder, \ie, have a 
larger binding energy. We performed illustrative scattering experiments 
with circular binaries and incoming stars with mass ratios of interest
to us in this study, using the scattering tools of STARLAB. The resulting 
distribution of the change in orbital radius $\Delta a/a$ is shown in 
Fig.~\ref{fig:delta}, and is seen to be highly asymmetric.

The long tail towards $\Delta a>0$ implies that the binary radius 
\emph{increases} in many scatterings. This does not of course contradict 
the above Heggie rule, since the increase of mass due to exchange (the 
mass of the incoming compact star, 1.4\Ms, is a factor $\approx 2.3$
times the mass of the outgoing low-mass star, 0.6\Ms) increases the 
binding energy by itself by the above factor. From these experiments, we 
see that the peak of the distribution corresponds to a shrinkage of the
binary by about 25 per cent. On the other hand, the \emph{average} 
change in binary radius, calculated from the above distribution, is much 
closer to zero due to the above long tail of the distribution on the  
$\Delta a>0$ side, so that we can take $a\approx a^\prime$ for our 
purposes here without much error. 

The total Maxwellian-averaged rate of formation of compact binary by 
this type of exchange (ex1) in the GC core is then:
\begin{equation}
r_{ex1}(a)=\frac{4}{3}\pi r_c^3 k_X \rho^2 f_b(a) \langle\sigma_{ex1}
(a)v\rangle=\sqrt{\frac{8\pi^3}{3}}k_Xf_b(a)\Gamma GM_{tot}a
\overline\sigma(m_c,m_X)
\label{eq:r_ex1}
\end{equation}
Here, $f_b(a)$ is the distribution function of the orbital radii of 
the primordial stellar binaries in the cluster core. For primordial 
binaries, we can take the widely-used distribution $f_b(a)\propto 1/a$ 
(\ie, a uniform distribution in $\ln a$) \citep{k78}, with a lower 
bound at $a\approx 1.2\Rs$, corresponding to the smallest possible 
radius for a binary of two $0.6\Ms$ main-sequence stars. The ex1 rate 
is shown in Fig.~\ref{fig:rates}.

\subsection{Binary destruction processes} \label{destroy}

A compact binary can be destroyed by two major processes. First, an 
encounter with a star which has a relative speed higher than an 
appropriate critical speed \citep{hb83} can lead to its dissociation 
(dss). Second, in an exchange encounter (ex2) of this binary with a 
compact star, the latter can replace the low-mass companion in the 
binary, forming a double compact-star binary consisting of two neutron 
stars, two white dwarfs, or a neutron star and a white dwarf, all with
masses $m_X\approx 1.4\Ms$. This, in effect, destroys the binary as an 
X-ray source (as accretion is not possible in such a system), and so 
takes it out of our reckoning in this study.  This is so because such a
system is not an X-ray source, and it is essentially impossible for one 
of the compact stars in such a system to be exchanged with an ordinary 
star in a subsequent exchange encounter, since $m_f=0.6\Ms$ is much 
lighter than $m_X=1.4\Ms$. The total destruction rate D(a) \emph{per 
binary} is thus the sum of the above dissociation and exchange rates:
\begin{equation}
D(a)=r_{ex2}(a) + r_{dss}(a)
\label{eq:d_tot}
\end{equation}
We now discuss the rates of these two processes.

\subsubsection{Dissociation}\label{dssdestroy}

To estimate the dissociation rate of compact binaries, we use the 
results of scattering experiments of \citet{hb83}. The 
Maxwellian-averaged dissociation rate (dss) \emph{per compact binary} 
is then given by
\begin{equation}
r_{dss}(a)=k_X \rho \langle\sigma_{dss}(a)v\rangle
\label{eq:r_dss}
\end{equation}
From \citet{hb83}, we adopt
\begin{equation}
\langle\sigma_{dss}(a)v\rangle=
\frac{32\pi}{27}\sqrt{\frac{6}{\pi}}v_ca^2
\exp\left(-\frac{3}{2}\frac{v_{crit}^2}{v_c^2}\right).
\label{eq:s_dssv}
\end{equation}
a relation which was obtained by these authors by fitting the results 
of their scattering experiments with analytical models. Here, 
$v_{crit}$ is the threshold relative velocity for ionization (see 
Sec.~\ref{destroy}), given by:
\begin{equation}
v_{crit}^2=\frac{Gm_X(2m_c+m_X)}{m_c+m_X}\frac{1}{a}.
\label{eq:vcrit}
\end{equation} 
As these authors pointed out, Eqn.~(\ref{eq:s_dssv}) is an asymptotic 
form, which works well only for significantly hard binaries, \ie, 
those with $v_c<<v_{crit}$. This condition is of course satisfied for 
the compact binaries that we are interested in here. 

We show in Fig.~\ref{fig:rates} the above dissociation rate, whose 
essential variation with $a$ is seen by combining Eqs.~(\ref{eq:s_dssv}) 
and (\ref{eq:vcrit}), which yields the form $r_{dss}(a)\propto a^2
\exp(-a_c/a)$, where $a_c$ is a constant. Thus, the dissociation rate is
quite negligible for $a\ll a_c$, reflecting the fact that it is essentially
impossible to dissociate very hard binaries. As $a$ increases, the rate
rises extremely sharply at first (the initial rise is determined by the 
exponential), and eventually scales as $a^2$ for $a\gg a_c$.  

\subsubsection{Destruction by exchange}\label{exdestroy}

By arguments similar to those given in Sec.~\ref{exform}, we arrive at a
Maxwellian-averaged rate of this type of exchange (ex2)  \emph{per 
compact binary} which is:
\begin{equation}
r_{ex2}(a)=k_X \rho\langle\sigma_{ex2}(a)v\rangle
=\sqrt{\frac{3\pi}{2}}k_X\gamma GM_{tot}a\overline\sigma(m_c,m_X),
\label{eq:r_ex2}
\end{equation}
and which is also shown in Fig.~\ref{fig:rates}. This rate scales with $a$
simply as $r_{ex2}(a)\propto a$. Note the different magnifications used 
for different curves in Fig.~\ref{fig:rates} in order to make all of them 
clearly visible. Of the two destruction processes, $r_{ex2}$ dominates 
completely at all orbital radii of interest in our study (reflecting the fact
that dynamically-formed binaries in GC cores are so hard that they 
cannot be dissociated or ``ionized'' by further encounters in that GC 
core), but the fast-rising $r_{dss}$ eventually overtakes it at $a\approx 
1000\Rs$, corresponding to very soft binaries.

\subsection{The numerical method}\label{num}

Equation~(\ref{eq:Evol}) for the evolution of compact binary populations is 
a partial differential equation (PDE) of hyperbolic type, with similarities to 
wave equations, as pointed out earlier. We solved this equation using a 
Lax-Wendorff scheme \citep{pr1992}. This involves dividing the range of 
$a$ and $t$ in a discrete mesh ($a_j,t_N$) of constant space intervals 
($\Delta a$) and time intervals ($\Delta t$). The PDE is then discretised into 
a set of linear difference equations over this mesh, which is solved 
numerically.

We denote by $n_j^N$ the value of $n$ at the $N$th time step and the $j$th 
point in $a$. Discretisation of Eqn.~(\ref{eq:Evol}) according to the 
Lax-Wendorff scheme is a two-step process:
\begin{equation}  
\begin{array}{l}
{\rm~Half~step:}\\
n_{j+1/2}^{N+1/2}={1\over2}\left(n_{j+1}^N+n_j^N\right)
+\left[R(a_{j+1/2}) - D(a_{j+1/2})\left(\frac{n_{j+1}^N+n_j^N}{2}
\right)\right]{\Delta t\over 2} \\
-\frac{f(a_{j+1/2})\Delta t}{2\Delta a}(n_{j+1}^N-n_j^N)\\ 
               \\
{\rm~Full~step:}\\
n_j^{N+1} = n_j^N + \left(R(a_j)-D(a_j)n_j^N\right)\Delta t\\
-\frac{f(a_j)\Delta t}{\Delta a}\left(n_{j+1/2}^{N+1/2}-n_{j-1/2}^{N+1/2}
\right)
\end{array}
\label{eq:lw}
\end{equation}
For a chosen mesh-interval $\Delta a$, Eqn.~(\ref{eq:lw}) will be numerically
stable only if the time-step $\Delta t$ is chosen to be small enough that it 
obeys the \emph{Courant condition} \citep{pr1992} throughout the mesh:
\begin{equation}
\Delta t=\eta\frac{\Delta a}{f_{max}},\quad\quad \eta<1
\label{eq:cc}
\end{equation}   
where $f_{max}$ is the maximum value of $f(a)$ within the $a$-range of the 
mesh.

We chose Lax-Wendorff scheme among the various existing schemes for 
solving hyperbolic PDEs primarily because it appears to be the only explicit 
method that does not have any significant numerical dissipation 
(\citealt{pr1992}, and references therein) and is at the same time numerically 
stable, provided that the time step is chosen according to the Courant 
condition. This point is important, since numerical dissipation can 
significantly affect the computed evolution of $n(a,t)$ and the X-ray binary 
population, as we observed while trying other methods, \eg, the so-called 
staggered-leapfrog method. Other instabilities, \eg, the mesh-drifting 
instability \citep{pr1992}, also appeared to be insignificant in the method 
we chose.

\section{Results}\label{results}

\subsection{Evolution of compact-binary distribution}\label{aevol}

\begin{figure}
\plotone{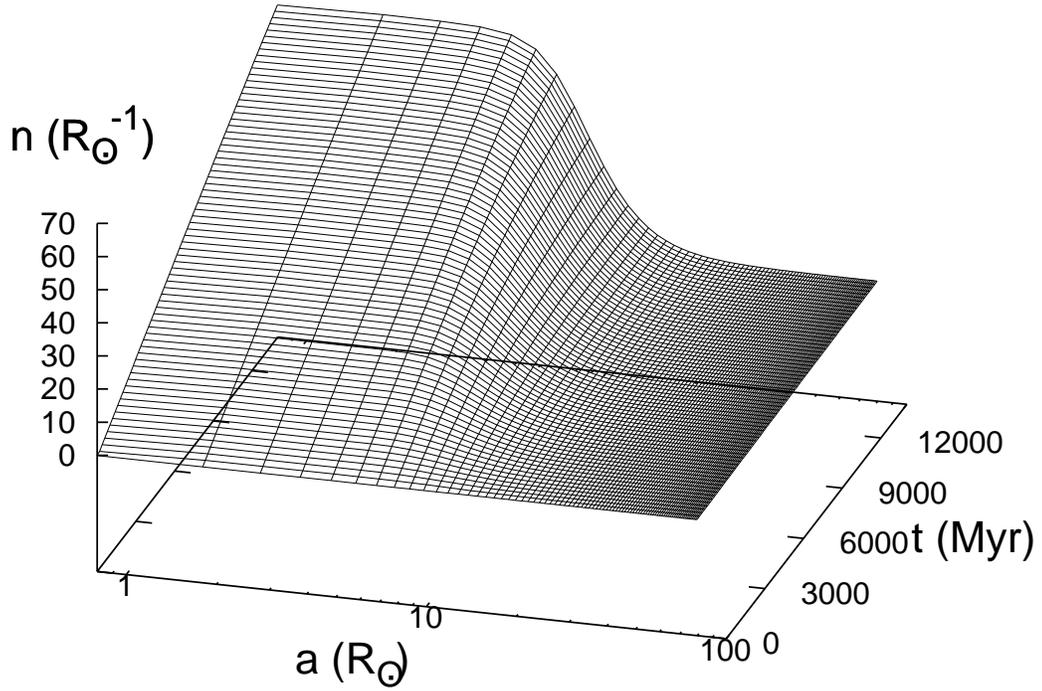}
\caption{Three-dimensional surface $n(a,t)$ describing the model 
evolution of population-distribution function of compact binaries 
for GC parameters $\rho=6.4\times10^4{\rm ~}\Ms{\rm ~pc}^{-3}$, 
$r_c=0.5{\rm ~pc}$, $v_c=11.6{\rm ~km}{\rm ~sec}^{-1}$ (roughly
corresponding to 47 Tuc). The lines on the surface represent only 
samples from the set of computed points, the computation having
been done over a much finer grid.}
\label{fig:ev3d}
\end{figure}

A typical result from our computed evolution of the compact-binary 
distribution function $n(a,t)$ is shown in Fig.~\ref{fig:ev3d}, wherein the 
surface $n(a,t)$ is explicitly displayed in three dimensions. The GC 
parameters chosen for this run were $\rho=6.4\times10^4{\rm~}\Ms
{\rm~pc}^{-3}$, $r_c=0.5{\rm~pc}$ and $v_c=11.6{\rm~km}{\rm~sec}
^{-1}$, similar to those of the well-known Galactic cluster  47 Tuc. 
The distribution function is seen to evolve as a smooth surface, with the 
compact binary population growing predominantly at shorter radii 
($a<10\Rs$, say). We start with a small number of binaries at $t=0$ 
following various distributions, and find that the distribution at large 
times $\sim$ Gyr is quite independent of these initial conditions, being 
determined entirely by the dynamical processes of formation and 
destruction, and by the various hardening processes detailed earlier.
Note that, since the point of Roche lobe contact corresponds to  $a 
\approx 2\Rs$ in our study, as explained earlier, that part of the 
distribution which is shortward of this radius corresponds to XBs, while 
that part longward of it corresponds to PXBs.

\begin{figure}
\plotone{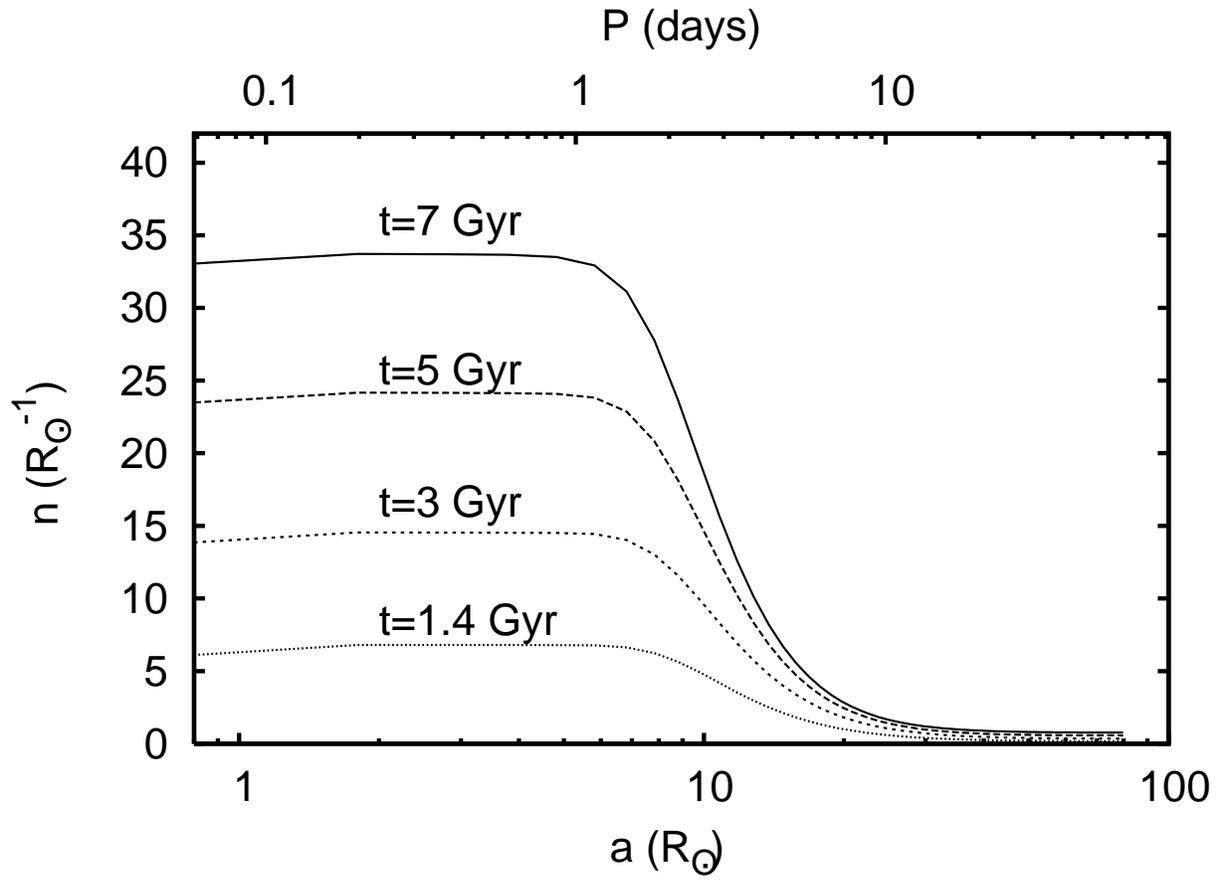}
\caption{Time slices, \ie, $n(a)$ at specified times $t$, for the 
evolution $n(a,t)$ shown in Fig.~\ref{fig:ev3d}. Along abscissa, 
both orbital radius $a$ and orbital period $P$ scales are shown 
for convenience.}
\label{fig:msnap}
\end{figure}

\begin{figure}
\plotone{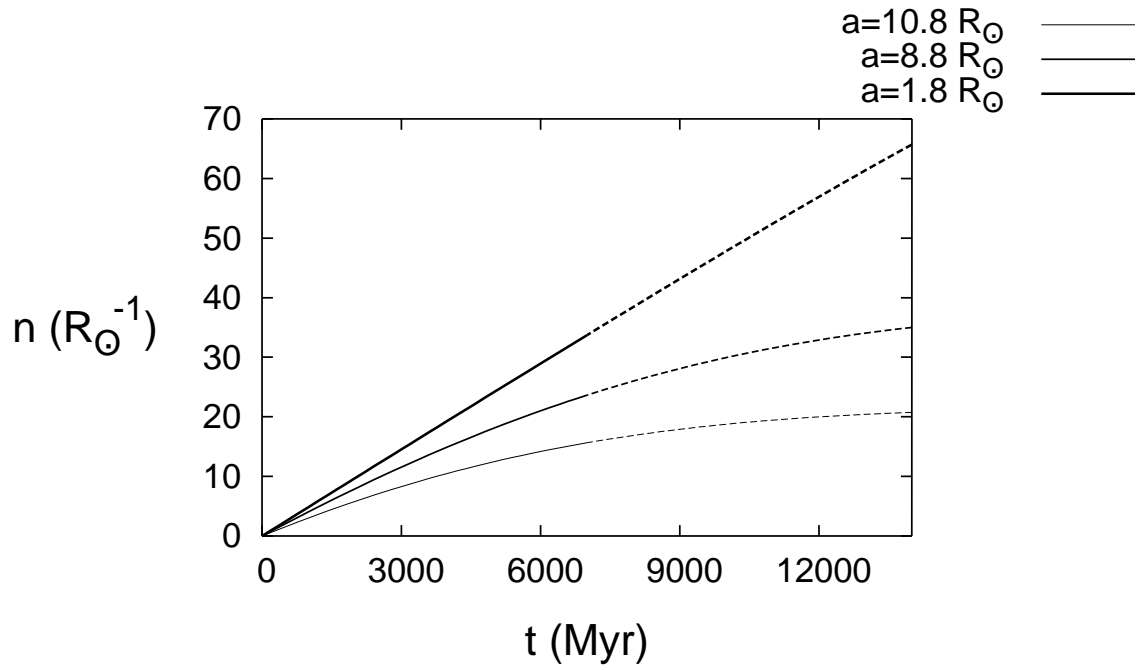}
\caption{Radial slices, \ie, $n(t)$ at specified orbital radii $a$, 
for the evolution $n(a,t)$ shown in Fig.~\ref{fig:ev3d}.}
\label{fig:tvol}
\end{figure}

To further clarify the nature of this evolution, slices through the 
above surface at various points along time axis and $a$-axis are shown 
in Figs.~\ref{fig:msnap} and \ref{fig:tvol}, in the former figure the 
abscissa being also marked in terms of the orbital period $P$, readily  
calculable in terms of $a$ and the stellar masses with the aid of 
Kepler's third law, assuming conservative mass transfer during the XB
phase. Figure \ref{fig:msnap} shows that $n(a)$ increases with time, 
roughly preserving its profile for $t>1.5$ Gyr or so. This profile 
consists of a roughly uniform distribution in for short orbital radii, 
$a\le 6\Rs$, say, corresponding to $P\le 1^d$ roughly, and a sharp 
fall-off at larger radii and orbital periods. Figure \ref{fig:tvol} 
shows that $n(a)$ at a given $a$ increases with time and approaches 
saturation on a timescale $6-12$ Gyr or so, this timescale being 
longer at at smaller values of $a$.

Figures \ref{fig:msnap} and \ref{fig:tvol} suggest that a regime of 
roughly self-similar evolution may be occurring in our model binary 
population at times beyond 1 Gyr or so, in the following way. An
asymptotic profile of $n(a)$ is established on the timescale of a 1
Gyr or so, which thereafter evolves roughly self-similarly towards 
a saturation strength on a timescale $\sim 6-12$ Gyr or so. We shall
discuss the origins of such behavior in detail elsewhere, since, as
explained in Sec.~\ref{apply}, our model of orbital evolution 
requires additional ingredients before it can be compared with     
observations of X-ray binaries. However, the following qualitative 
remarks are appropriate here. 

First, the origins of the establishment 
of the above self-similar profile in a Gyr or so (independent of the 
initial distribution we start from) clearly lie in the two terms 
that describe binary formation and hardening on the right-hand side 
of Eq.~(\ref{eq:Evol}), namely, $R(a)$ and $\frac{\partial n}
{\partial a}f(a)$ respectively. The latter term can be written 
qualitatively in the form $n/\tau_h$, where $\tau_h$ is the overall
hardening timescale, which is well-known to be of the order of a
Gyr or so (see BG06 and references therein). This timescale, which
is also that on which a given binary passes from the large-$a$ end
of the distribution shown in Fig.~\ref{fig:ev3d} to the small-$a$   
end, is obviously the timescale that establishes the above profile. 
The shape of this profile, as detailed above, seems related to those
of the tidal-capture rate (see Fig.~\ref{fig:rates}) and the 
hardening rate (see Fig.~\ref{fig:adot}). In particular, note that
the former rate is roughly constant over $a_{min}\le a\le 5\Rs$,
and the latter roughly so for $a_{pm}\le a\le 2\Rs$. 

Second, the subsequent, roughly self-similar evolution of the above
profile occurs on a (longer) timescale $\tau_s$ whose origins 
clearly lie in the binary destruction term on the right-hand side 
of Eq.~(\ref{eq:Evol}), namely, $nD(a)$, since this term can be 
cast in the qualitative form $n/\tau_s$, where $\tau_s$ is the    
saturation time $\sim 6-12$ Gyr. Whereas the earlier term  
$n/\tau_h$ describes the passage or ``current'' of binaries 
through the distribution, as described earlier, the term $n/\tau_s$
becomes important as $n$ increases, preventing $n$ from becoming
arbitrarily large by enforcing saturation at the point where the 
rates of formation and destruction balance. As $D(a)$ scales with
$a$, as shown above, and $\tau_s=1/D(a)$, we expect saturation to
occur at earlier times at larger radii, as seen in 
Fig.~\ref{fig:tvol}. 

\subsection{Number of X-ray binaries in globular 
clusters}\label{xbnumber}

The total number of X-ray binaries $N_{XB}$ in a GC at any time 
can be computed directly from our approach by integrating $n(a,t)$ 
over the range of $a$ relevant for XBs, \viz, $a_{pm}\le a\le a_L$, 
where $a_{pm}$ is the value of $a$ corresponding to the period
minimum $P\approx 80$ minutes, and $a_L$ is the value of $a$ at the 
first Roche lobe contact and onset of mass transfer, as explained 
earlier. We have: 
\begin{equation}
N_{XB}(t)=\int_{a_{pm}}^{a_L}n(a,t)da
\label{eq:nxb}
\end{equation} 

Taking an evolutionary time $\sim 8$ Gyr as representative, we can 
therefore determine $N_{XB}$ at this point in time, and study its 
dependence on the Verbunt parameters $\Gamma$ and $\gamma$ that 
describe the essential dynamical properties of globular clusters in 
this context, as explained earlier. By doing so, we can attempt to 
make qualitative contact with the systematics of those recent 
observations of X-ray binaries in globular clusters which we have 
described earlier \citep{pool03}. To this end, we computed values of 
$N_{XB}$ over a rectangular grid spanning over $\gamma=1-10^6$ and 
$\Gamma=10^3-10^8$. (Of course, not all the points on the grid would 
be directly relevant for comparison with observation, since the 
observed globular clusters lie only along a diagonal patch on this 
grid, as shown in Fig.~\ref{fig1}. However, in this introductory
study, we wished to establish the theoretically expected trends of 
variation with $\Gamma$ and $\gamma$, and so performed computations 
of $N_{XB}$ over the entire rectangular grid)

For a specified grid point, \ie, a pair of values of the Verbunt 
parameters, we obtained values of $\rho$, $r_c$ and $v_c$ with the 
aid of Eqs.~(\ref{eq:Gamma}), (\ref{eq:gamma}) and the virialization 
condition:
\begin{equation}
v_c \propto \rho^{1/2}r_c
\label{eq:vt}
\end{equation}
which were used for the computation at this grid point. We chose this 
prescription for the sake of definiteness, because values of $v_c$ 
are not known, in general, at a computational grid point, without 
which a pair of Verbunt parameters cannot specify all three variables 
$\rho$, $r_c$ and $v_c$. This also introduced a certain uniformity 
of treatment of all grid points, which, we thought, would clarify the 
theoretically expected trends. On the other hand, this did lead to a 
feature at high values of $\Gamma$ and low values of $\gamma$,\ie, in 
that part of the grid which is completely devoid of observed globular 
clusters at this time (and which, in fact, may \emph{actually} contain 
no clusters, because such combinations of $\Gamma$ and $\gamma$ may 
not be possible in nature), which appears unphysical, as we discuss 
below. Observationally, we know, of course, that some clusters appear
fairly virialized and some do not, but any spread in $v_c$ applied 
over the grid points would have been arbitrary, and would have led to 
a scatter, masking the systematic theoretical trends without purpose. 
Finally, throughout these computations, we used representative values 
for (a) the primordial binary fraction $k_b$, namely, 10 percent, and 
(b) compact star fraction $k_X$, namely, 5 percent.  

\begin{figure}
\plotone{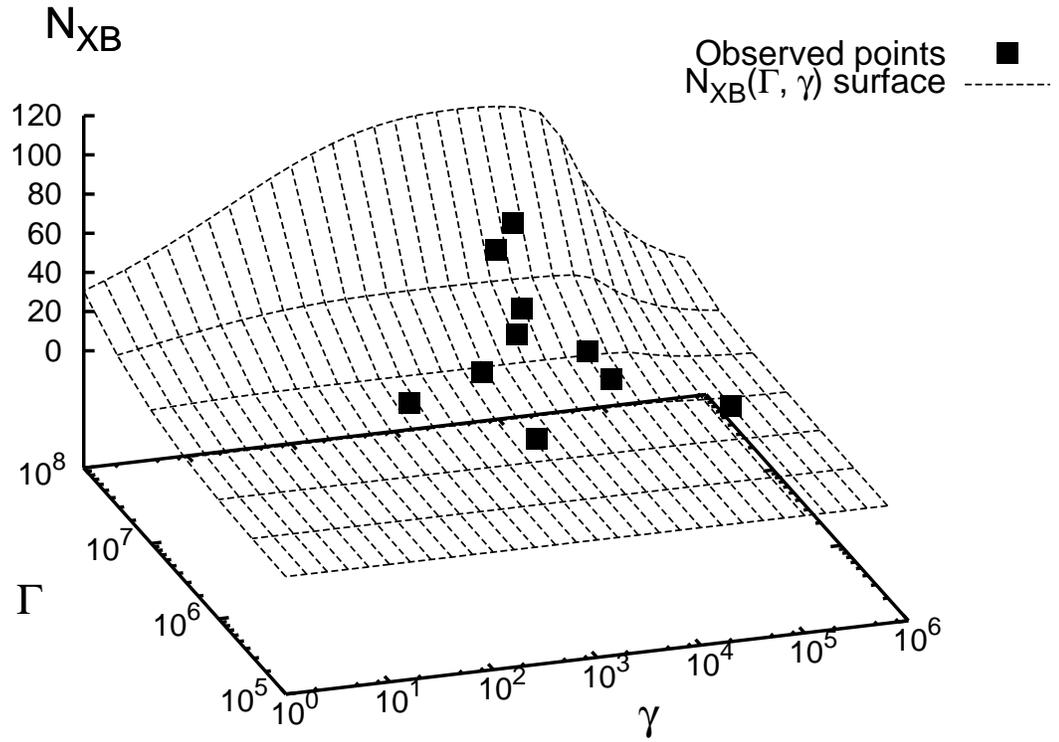}
\caption{Computed $N_{XB}(\Gamma,\gamma)$ surface. Overplotted are 
the positions of the globular clusters with significant numbers of 
X-ray sources (filled squares) from Fig.~\ref{fig1}.}
\label{fig:NXB}
\end{figure}

Figure~\ref{fig:NXB} shows the computed surface $N_{XB}(\gamma, 
\Gamma)$. There appears to be a ``fold'' in this surface, in a 
direction roughly parallel to the $\Gamma$ axis, located around 
$\gamma=3\times10^3$.  From this fold, if we go towards higher values 
of $\gamma$, then, for any given value of $\Gamma$, $N_{XB}$ decreases 
with increasing $\gamma$. This is a signature of the compact-binary 
destruction processes detailed in the previous section, whose strengths 
increase with increasing $\gamma$. Thus, the above value of $\gamma$
corresponding to the fold seems to be a good estimate of the threshold 
above which these destruction processes dominate. At constant $\gamma$, 
the variation with $\Gamma$ is quite straightforward: $N_{XB}$ simply 
increases monotonically with increasing $\Gamma$, reflecting the fact 
that the formation rates of compact binaries, as described in the 
previous section, increase with increasing $\Gamma$.  

\begin{figure}
\plotone{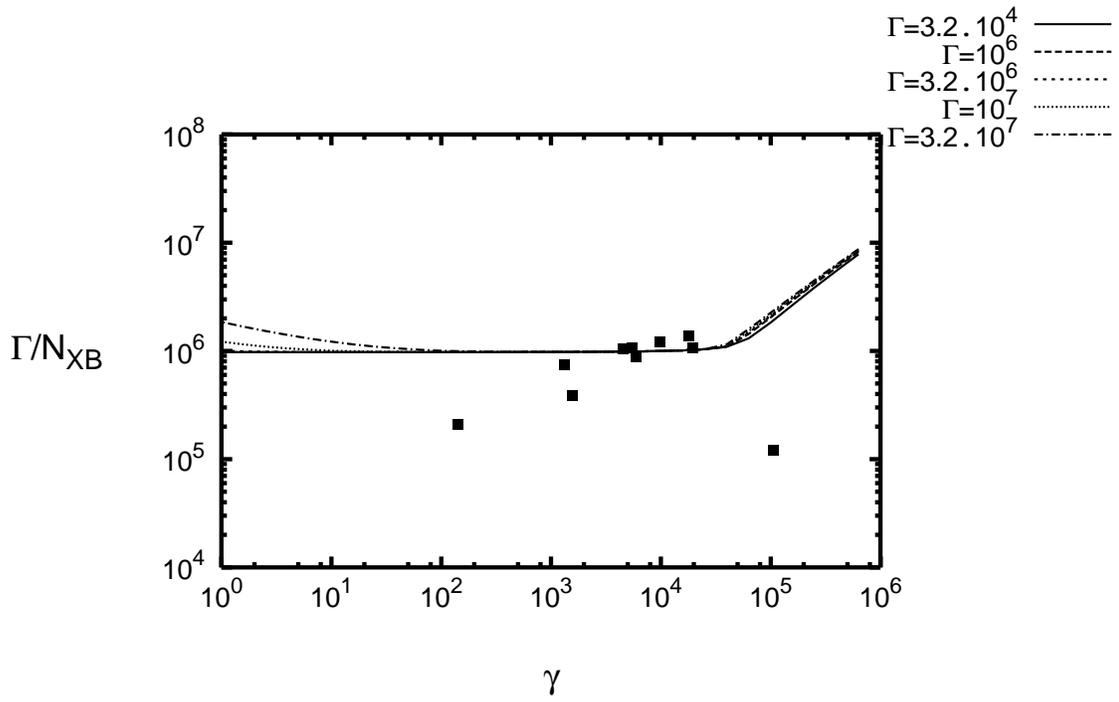}
\caption{Computed $\Gamma/N_{XB}$ as a function of $\gamma$, 
showing scaling (see text). Computed curves for various values of 
$\Gamma$ are closely bunched, as indicated. Overplotted are the 
positions of the globular clusters with significant numbers of 
X-ray sources (filled squares) from Fig.~\ref{fig1}.}
\label{fig:toylike}
\end{figure}

To further clarify these trends, and to facilitate comparison with those 
obtained from the ``toy'' model of BG06, we display in Fig.~\ref{fig:toylike} 
$\Gamma/N_{XB}$ vs. $\gamma$, as was done in that reference. The 
motivation is as follows. It was shown in BG06 that the toy model of these 
authors leads to the scaling that $\Gamma/N_{XB}$ was a function of 
$\gamma$ alone, which was a monotonically increasing function of 
$\gamma$, for which the toy model gave a very simple, analytic form. 
Our purpose in Fig.~\ref{fig:toylike} is to see how much of this scaling 
survives the scrutiny of a more detailed model, such as presented here.
As the figure shows, this scaling does carry over approximately, although 
some details are different. $\Gamma/N_{XB}$ is still almost a function of 
$\gamma$ alone (except at the very highest values of $\Gamma$), showing 
that this scaling $N_{XB}\propto\Gamma g(\gamma)$ of the toy model 
carries over approximately to more detailed ones, thereby giving an indication 
of the basic ways in which dynamical binary formation and destruction 
processes work. The above ``universal'' function $g(\gamma)$ of $\gamma$ 
is, except for a feature at low values of $\gamma$ which we discuss
below, still a monotonically increasing one, reflecting the increasing strength
of dynamical binary-destruction processes with increasing $\gamma$. 
However, the shape of the function is different in detail now, as may have 
been expected.   

We now discuss the low-$\gamma$ feature referred to above: at the lowest 
values of $\gamma$, $\Gamma/N_{XB}$ seems to rise again, reflecting an 
apparent drop in $N_{XB}$. This is difficult to understand, since 
binary-destruction effects are negligible at these values of $\gamma$. 
Actually, this is an artifact of the way in which we fixed the essential 
cluster parameters $\rho$, $r_c$ and $v_c$ from specified values of the 
Verbunt parameters for the computational grid (as explained above), which 
can be seen as follows. With the assumption of virialization, as done for 
this purpose, the velocity dispersion $v_c$ can be expressed in terms of 
the Verbunt parameters in a manner analogous to that used in 
Eq.~(\ref{eq:invert}), the result being $v_c\propto\Gamma^{2/5}
\gamma^{-1/5}$. This relates $v_c$ to $\gamma$, so that the latter 
influences the Maxwellian-averaging process involved in the calculation 
of the tidal capture cross-section described in Sec.~\ref{tcform}, since 
the parameter $\beta\equiv 3/(2v_c^2)$ of the Maxwellian then scales 
as $\beta\propto\Gamma^{-4/5}\gamma^{2/5}$. At small values of 
$\gamma$, $\beta$ becomes small, which reduces the tidal-capture rate, 
as Eq.~(\ref{eq:r_tc2}) readily shows. This is completely unphysical, of 
course, since $\gamma$ has nothing to do physically with the tidal 
capture rate. Rather, it is an artifact produced by the way we 
(artificially) related $v_c$ to $\gamma$ for computational convenience.
Accordingly, we ignore this low-$\gamma$ feature in all further 
considerations.   

\section{Comparison with Observation}\label{cfobs}

\subsection{Applicability of our study}\label{apply}

Before attempting to compare our results with observations, we review
in brief some essential ingredients of our model study at this stage,
so as to clarify which of our results can be so compared, and which
need inclusion of further components before this can be meaningfully
done. A major ingredient that is incomplete at this stage is our 
description of the orbital evolution of the binary, since it neglects 
nuclear evolution of the low-mass companion star altogether. While
this may not be very unreasonable for CV systems or for short-period 
LMXBs with orbital periods between $\sim 10$ hours and the above 
period minimum of $\sim 80$ minutes, it is completely inadequate for 
other LMXBs, where the stellar evolution of the companion
plays a crucial role, which has been studied by many authors. In
particular, recent studies by \citet{prp02} and \citet{prp03} have
demonstrated the large range of possibilities covered by such
evolution with realistic stellar evolutionary codes, performing a 
Monte Carlo binary population synthesis study in the latter reference
with the aid of the library of evolutionary sequences described in
the former. We plan to include stellar evolutionary effects in a 
subsequent work of the series and are assessing various methods of 
doing so. One possibility is to start with a semi-analytic scheme 
along the lines of the SeBa model as described in \citet{z297}, and
to continue with a semi-analytic approximation to a more elaborate 
library of evolutionary sequences, such as described above. 

Since most of the XBs in the Galactic GC data of \citet{pool03} are
CVs, our scheme should be able to describe the overall properties of 
these XB populations reasonably well. Even so, we shall make no
attempt here to compare our results on orbital period distribution 
with the observed CV distribution, since the CVs in the latter 
distribution are almost exclusively from outside globular clusters,
where dynamical formation is not relevant. We have here recorded the
orbital-period distribution that comes from our computations (at 
this stage) only as a natural intermediate step. It can perhaps be   
compared with observation when the orbital-period distribution of
CVs \emph{in GCs} becomes observationally established. For LMXBs,
where the observed orbital-period distribution at this time also 
consists overwhelmingly of those outside GCs, there is of course no
question of comparison at this stage, for the reasons given above.
Thus, our main aim here is to put in the observational context our 
results on the numerical properties of XB populations in GCs in
relation to the GC parameters.   

\subsection{Ultracompact X-ray binaries}\label{ucxb}

In recent years, a subset of LMXBs in GCs, in the Milky Way and 
possibly also in elliptical galaxies, have received much attention 
because of (a) their high, persistent brightness ($L_x\sim 10^{36}-
10^{39}$ erg s$^{-1}$), which would make them dominate the high end 
of the luminosity functions of X-ray binaries in ellipticals  
\citep{bd04} and (b) their very close 
orbits with $P<1$ hr or so, sometimes as short as $P\sim 10$ minutes, 
the classic example being the 11 min binary 4U 1820-30 the Galactic 
cluster NGC 6624. These are the ultracompact X-ray binaries (henceforth 
UCXBs), which are thought to consist of neutron stars in ultracompact
orbits with very low-mass degenerate dwarf companions ($m_c\sim
(0.06-0.2)\Ms$) as mass donors. The evolutionary origin of UCXBs is
of much current interest, and proposals for such origins include 
(a) direct collisions between red giants and neutron stars in GC 
cores, as a consequence of which the red-giant envelope can either
be promptly disrupted \citep{i.et.al05} or be expelled more slowly 
in a common-envelope phase, and (b) usual LMXB evolution with the 
initial orbital period below the ``bifurcation period'' of about 18 
hrs \citep{prp02}. A natural point that arises, therefore, is about 
the role of UCXBs in our study, and the general importance of the 
above channels of formation in relation to the ones we have described 
above, which we now consider in brief.   

The key feature of UCXBs from the point of view of our study is that
the number of UCXBs $N_{UC}$ is a tiny fraction of the total number 
of XBs in a GC, and so of little importance as far as $N_{XB}$ is 
concerned. This is a general, robust feature, which follows from the 
basic point that the UCXBs are extremely short-lived because of their 
extreme brightness, so that $N_{UC}$ is small at any given epoch
despite their considerable birthrate. To see this in more detail,     
consider the UCXB birthrate of about one every $2\times 10^6$ year
per $10^7\Ms$ of the mass of a GC, as given by \citep{bd04}, which, 
together with their estimated lifetimes of $(3-10)\times 10^6$ years, 
yields an estimate of $N_{UC}\sim 1-5$ in a $10^7\Ms$ GC at any given 
time. Actually, the observed GCs in our galaxy have lower masses, in 
the range $\sim (10^5-10^6)\Ms$ \citep{i.et.al05}. Thus a Galactic
GC of $10^6\Ms$ like 47 Tuc will have $N_{UC}\sim 0.1-0.5$, 
remembering that the birthrate scales down appropriately with the
GC mass, but the lifetime remains the same. This is to be compared
with the observed number of XBs in 47 Tuc of 45 \citep{pool03}, 
which yields a fraction $N_{UC}/N_{XB}\sim 2\times 10^{-3}-1.1
\times 10^{-2}$. We can double-check this and put it on a systematic    
basis with the aid of Table 1 of \citet{i.et.al05}, wherein these 
authors have listed the minimum expected number of UCXBs in a number
of Galactic GCs, by combining this with the total number of observed  
XBs obtained from \citet{pool03} and other sources. For 47 Tuc, with 
0.23 UCXBs and 45 XBs, the ratio is $N_{UC}/N_{XB}\sim 5\times 
10^{-3}$, very similar to above, and those for other sources are
also similar. For example, Terzan 5 has a ratio $\sim 2\times 
10^{-3}$, and NGC 6652 has a ratio $\sim 8\times 10^{-4}$. 

We see from the above that UCXBs constitute such a tiny fraction of 
the total XB populations of Galactic GCs in terms of numbers that 
their effect is negligible for this work. However, in a study of the 
X-ray luminosity functions of GCs, their effect is expected to be      
crucial: if a GC contains even one UCXB, its luminosity may dominate
over the combined output of all other XBs. It is the extension of this
idea which has been used in recent years to argue that the luminosity
function of XBs in ellipticals may be dominated by UCXBs in their
GCs \citep{bd04}.

\subsection{X-ray source numbers in globular clusters}\label{xnumcfobs}

\begin{figure}
\plotone{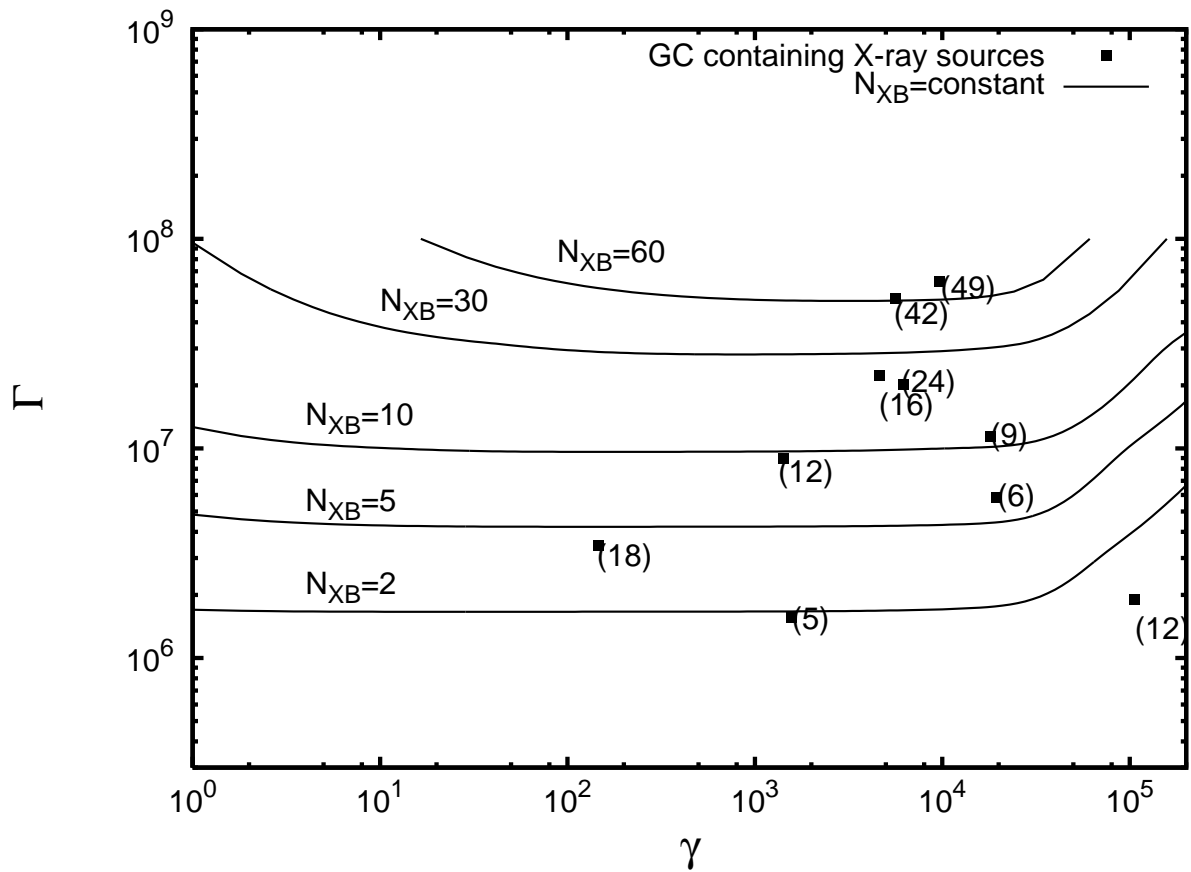}
\caption{Contours of constant $N_{XB}$ in the $\Gamma-\gamma$ 
(Verbunt parameters) plane. Overplotted are positions of Galactic 
globular clusters with significant numbers of X-ray sources detected 
in them (filled squares) from Fig.~\ref{fig1}. 
$N_{XB}$ for each cluster is indicated by the number in 
parenthesis next to its marked position. }
\label{fig:cont}
\end{figure}

The filled squares in Fig.~\ref{fig:NXB} represent globular clusters 
with significant numbers of X-ray binaries in them. These points 
generally lie near the surface in this three-dimensional plot, mostly 
in the vicinity of the fold described above. This is more clearly seen 
in the two-dimensional plot of Fig.~\ref{fig:toylike}, where the bulk 
of the observational points are indeed seen to be near the upward 
``knee'' of the computed curves. To facilitate comparison with 
observations further, we show in Fig.~\ref{fig:cont} contours of 
constant $N_{XB}$ in the $\Gamma-\gamma$ (Verbunt parameters) plane. 
Overplotted on these are the above observed clusters (filled sqaures), 
where the number in the parentheses next to each indicates the 
total number of X-ray binaries observed in it \citep{pool03}. The 
contours are seen to be qualitatively rather similar in shape to the 
curves in Fig.~\ref{fig:toylike}. The trend in the observed $N_{XB}$ 
values generally follows the contours, with one exception. This is most 
encouraging (also see BG06) for the construction of more detailed 
models, and indeed rather remarkable in view of the fact that no 
particular attempt has been made to fit the data at this stage.     

\section{Discussion}\label{discuss}

In this paper, we have explored the results of a Boltzmann study of 
the evolution of compact-binary populations in globular clusters in 
the continuous limit, and made preliminary contacts with observations 
of X-ray binaries in Galactic globular clusters. Our Boltzmann approach 
has built into it the rates of the essential dynamical processes that 
occur due to star-star and star-binary encounters in dense clusters, 
\viz, collisional hardening, binary formation by tidal capture and 
exchange, and binary destruction by dissociation and other 
mechanisms, as obtained by previous numerical studies of large numbers 
of such individual encounters. We stress that our Boltzmann scheme is 
not a Fokker-Planck one, wherein the cumulative effects of a large 
number of small changes in distant encounters is described as a slow 
diffusion in phase space. We can and do handle both small and large 
changes in the framework of the original Boltzmann visualization of 
motion through phase space (at a computational cost which is quite 
trivial compared to that required for N-body simulations).
Indeed, the continuous limit of collisional 
hardening used in this paper may be looked upon as an example of a slow 
diffusion in $a$-space, while some of the formation and destruction 
processes are examples of faster and more radical changes. Of course, 
all these processes are episodic in nature, and we are studying their 
continuous, probabilistic limit in this introductory paper. As already 
pointed out, Paper II will describe an explicit treatment of the 
stochasticity of these processes within the framework of stochastic PDEs,
which the Boltzmann equation becomes under such circumstances.    

\subsection{Conclusions}\label{conclude}

We find the indications from this preliminary study to be sufficiently 
encouraging to attempt several steps of improvement, most of which 
we have already indicated in the previous sections. To recapitulate 
briefly, we need to provide an appropriate description of the 
stochastic processes, which we do in Paper II. We need to introduce 
a mass function for the background stars in the globular cluster core, 
and handle non-circular orbits formed in the encounter processes. We 
need to assess the possible importance of binary-binary interactions 
in this problem, which we have neglected altogether so far. We need 
to include essential aspects of stellar evolution of the companion
in our orbital-evolution scheme, particularly for LMXBs. In a more 
ambitious vein, we need to consider the evolution of the stellar 
background representing the cluster core, which we do in Paper III. 
As the core collapses, the collapse stalls due to binary heating, and 
possible gravothermal oscillations occur, the core parameters $\rho$ 
and $r_c$ evolve appropriately, and so do the Verbunt parameters 
$\Gamma$ and $\gamma$. Binary-population evolution with such evolving 
GC parameters is an interesting problem in itself, even if we do not 
explicitly consider the back reaction of binary evolution on the 
evolution of its background. 

The scaling of $N_{XB}$ with the two Verbunt parameters we already 
found here seems to be among the basic building blocks of our 
understanding of how globular clusters cook up their gross 
overabundance of X-ray binaries through an interplay between dynamical 
formation and destruction. It remains to be seen if there are other
such building blocks which have not been investigated so far.  

\acknowledgments

It is a pleasure to thank H. M. Antia, D. Heggie, P. Hut, S. Portegies 
Zwart, and F. Verbunt for stimulating discussions and correspondence,
and the referee for many suggestions which considerably improved the 
paper.

%% Note that the style of the \bibitem labels (in []) is slightly
%% different from previous examples.  The natbib system solves a host
%% of citation expression problems, but it is necessary to clearly
%% delimit the year from the author name used in the citation.
%% See the natbib documentation for more details and options.

\clearpage

%% Use the figure environment and \plotone or \plottwo to include
%% figures and captions in your electronic submission.
%% To embed the sample graphics in
%% the file, uncomment the \plotone, \plottwo, and
%% \includegraphics commands
%%
%% If you need a layout that cannot be achieved with \plotone or
%% \plottwo, you can invoke the graphicx package directly with the
%% \includegraphics command or use \plotfiddle. For more information,
%% please see the tutorial on "Using Electronic Art with AASTeX" in the
%% documentation section at the AASTeX Web site,
%% http://www.journals.uchicago.edu/AAS/AASTeX.
%%
%% The examples below also include sample markup for submission of
%% supplemental electronic materials. As always, be sure to check
%% the instructions to authors for the journal you are submitting to
%% for specific submissions guidelines as they vary from
%% journal to journal.

%% This example uses \plotone to include an EPS file scaled to
%% 80% of its natural size with \epsscale. Its caption
%% has been written to indicate that additional figure parts will be
%% available in the electronic journal.

\clearpage

%%%%%%%%%%%%%%%%%%%%%%%%%%%%%%%%%%%%%%%
%%%%%%%%%%%%%%%%%%%%%%%%%%%%%%%%%%%%%%%

%%%%%%%%%%%%%%%%%%%%%%%%%%%%%%%%%%%%%%%

%%%%%%%%%%%%%%%%%%%%%%%%%%%%%%%%%%%%%%%%

%%%%%%%%%%%%%%%%%%%%%%%%%%%%%%%%%%%%%%%

%%%%%%%%%%%%%%%%%%%%%%%%%%%%%%%%%%%%

%%%%%%%%%%%%%%%%%%%%%%%%%%%%%%%%%%%%%%%

%%%%%%%%%%%%%%%%%%%%%%%%%%%%%%%%%%%%%%%%

%%%%%%%%%%%%%%%%%%%%%%%%%%%%%%%%%%%%%%%%

%%%%%%%%%%%%%%%%%%%%%%%%%%%%%%%%%%%%%%%

%%%%%%%%%%%%%%%%%%%%%%%%%%%%%%%%%%%%

%% The following command ends your manuscript. LaTeX will ignore any text
%% that appears after it.

\end{document}